\documentclass[12pt]{article}
\usepackage{graphicx}
\usepackage{mathptmx,mathrsfs}
\usepackage{amsfonts,amsmath,amssymb,amsthm}
\usepackage{indentfirst}
\usepackage{cite}
\usepackage{hyperref}
\usepackage{enumitem}

\numberwithin{equation}{section}
\begin{document}

\title{Functional quantization of Generalized Scalar Duffin-Kemmer-Petiau
Electrodynamics}
\author{R. Bufalo$^{1}$\thanks{%
rbufalo@ift.unesp.br}~, T.R. Cardoso$^{1}$\thanks{%
cardoso@ift.unesp.br}~, A.A. Nogueira$^{1}$\thanks{%
nogueira@ift.unesp.br}~, B.M. Pimentel$^{1}$\thanks{%
pimentel@ift.unesp.br} \\
\textit{{$^{1}${\small Instituto de F\'{\i}sica Te\'orica (IFT),
Universidade Estadual Paulista}}} \\
\textit{\small Rua Dr. Bento Teobaldo Ferraz 271, Bloco II Barra Funda, CEP
01140-070 S\~ao Paulo, SP, Brazil}\\
}
\maketitle
\date{}

\begin{abstract}
The main goal of this work is to study systematically the quantum aspects of
the interaction between scalar particles in the framework of Generalized
Scalar Duffin-Kemmer-Petiau Electrodynamics (GSDKP). For this purpose the
theory is quantized after a constraint analysis following Dirac's
methodology by determining the Hamiltonian transition amplitude. In
particular, the covariant transition amplitude is established in the
generalized non-mixing Lorenz gauge. The complete Green's functions are
obtained through functional methods and the theory's renormalizability is
also detailed presented. Next, the radiative corrections for the Green's
functions at $\alpha $-order are computed; and, as it turns out, an
unexpected $m_{P}$-dependent divergence on the DKP sector of the theory is
found. Furthermore, in order to show the effectiveness of the
renormalization procedure on the present theory, a diagrammatic discussion
on the photon self-energy and vertex part at $\alpha ^{2}$-order are
presented, where it is possible to observe contributions from the DKP
self-energy function, and then analyse whether or not this novel divergence
propagates to higher-order contributions. Lastly, an energy range where the
theory is well defined: $m^{2}\ll k^{2}<m_{p}^{2}$ was also found by
evaluating the effective coupling for the GSDKP.
\end{abstract}

\newpage

\section{Introduction}

The Duffin-Kemmer-Petiau (DKP) equation is a first-order relativistic theory
for the description of spin $0$ and spin $1$ bosons with a similar form as
the Dirac equation\footnote{%
The historical development of this theory, among others, can be found in
\cite{Histdkp,histrev}.}. Substantiated on Imbert's experiments \cite{imbert}%
, which suggested strong classical and quantum contradictions for
longitudinal plane waves displacement, de Broglie states that a possible
non-zero rest mass to photons would be the right interpretation for that
phenomenon \cite{broglie}. In fact, de Broglie also suggested that the photon
should be formed by the combination of two leptons and such a combination
should be responsible for assign a mass to photon. Driven by this idea and
with a deep knowledge on the algebraic structure of Dirac's equation
(relativistic equation for spin $1/2$ particle), de Broglie begins his
search of a first-order equation in the hope of obtaining an equation for a
massive particle of spin $1$, his massive photon \cite{sidarta}.

Petiau was the first to obtain the matrix algebra of DKP \cite{petiau}%
\footnote{%
Forsooth, G\'{e}h\'{e}niau decomposed Petiau's sixteen-dimensional algebra
in terms of irreducible representations of ten dimensions (representing
particles of spin $1$), five dimensions (representing particles of spin $0$%
), and a trivial representation without physical meaning of one dimension
\cite{tadinho}.}. Simultaneously and completely unaware to the work of
Petiau, Kemmer wrote the second-order Proca equations and the
Klein-Gordon-Fock (KGF) equation as a set of coupled first-order equations.
Kemmer then conjectures about the existence of a matrix form describing this
system of coupled equations, on which irreducible representations
representing particles of spin $0$ (scalar particles) and spin $1$
(vectorial particles). Duffin develops the desired algebra for Kemmer's
theory \cite{nieto, duffin,kemmer}.

The DKP formalism allows an unified treatment of the scalar and vector fields%
\footnote{%
This formalism can be extended to describe non-Abelian and gravitational
fields \cite{Okubo}.} and the wealth of couplings in the DKP formalism made
{}{}the theory initially well received; in fact, due to its unique algebraic
structure this formalism enjoys a plenty of couplings incapable of being
expressed in the theories of KGF and Proca \cite{Umez, corson, takahashi,
scott, refs}. However, the equivalence of DKP and KGF in the free and
minimal electromagnetic coupling cases \cite{pims, fai, fain, bra}, both in
classical and quantum pictures, led to a decreased interest in DKP theory.
Although the KGF formalism is apparently simpler when compared to the
algebraic treatment of the DKP theory in a classical picture, this point of
view changes dramatically in a quantum picture: the similar form between the
DKP and Dirac Lagrangian expressions allows a very simplified mechanism to
study scalar phenomena, once the mimetism with Dirac theory can be used to
understanding of the physical meaning of all the quantities obtained from
the DKP theory \cite{kinoshita,aqui}.

In the last years the DKP theory has been studied on QCD at large and short
distances by Gribov \cite{Gri}, in the scattering $K^{+}$-nucleus \cite%
{Kmais}, covariant Hamiltonian dynamics \cite{Kana}, in generalization to
curved space-time \cite{Red}, in a five-dimensional Galilean covariance \cite%
{5d}, in the context of classical gauge invariance \cite{rem}, in the
Epstein--Glaser causal method \cite{lucausal} and so on.

Although the extensive research concerning the DKP theory within the
framework of gauge theory, it is desirable to consider its interaction with
distinct gauge fields. It is a well-known fact that Maxwell's
electrodynamics is considered as being one of the most successful physical
theories; despite that, the research in pursuing gauge-invariant
alternatives extensions in order to supplement it is an ongoing subject of
study \cite{ref15,ref76}. Among the several features that these variants are
endowed, the main difference between these theories is due to the
nonlinearity of the field equation, e.g. Born-Infeld and Euler-Heisenberg
Lagrangians, while the linearity of the field equation with higher-order
derivatives, e.g. Bopp-Podolsky \cite{Bop1,Pod1}.

It is a well-known fact that higher-order derivative (HD) theories have \cite%
{Ostro, ref15}, in the light of effective field theory \cite{Weinb}, better
renormalization properties than the conventional ones\footnote{%
This idea is successful in the case of the attempt to quantize gravity,
where the (non-renormalizable) Einstein action is supplied by terms
containing higher powers of curvature leading to a renormalizable \cite{ref4}%
. Also, a new impetus in exploring appealing quantum theories such as $%
f\left( R\right) $-gravity \cite{ref6}}. One of the most interesting
contributions to show the effectiveness of the HD terms in field theory is
the Bopp-Podolsky electrodynamics, a generalization of the Maxwell
electromagnetic field \footnote{%
A non-Abelian version of the Bopp-Podolsky electrodynamics was studied and
deeply analyzed in \cite{ref3,ref38}.}. Moreover, the Ref.~\cite{Ut} showed
that the Podolsky Lagrangian is the only linear generalization of Maxwell
electrodynamics that preserves invariance under $U\left( 1\right) $. An
important feature concerning the Bopp-Podolsky electrodynamics is the gauge
fixing, i.e. on how to fix the correct physical degrees of freedom, since
the usual Lorentz condition is not suitable. It was shown in the Ref.~\cite%
{galvao} that the natural condition in the Bopp-Podolsky electrodynamics is
the generalized Lorentz condition, $\Omega \lbrack A]=\left( 1+a^{2}\square
\right) \partial ^{\mu }A_{\mu }$. However, there are alternative gauge
conditions that allow the same identification, in particular, and in order
to preserve the order of the field equation, the so-called non-mixing gauge
term \cite{LW, Daniel}, $\Omega \lbrack A]=\left( 1+a^{2}\square \right) ^{%
\frac{1}{2}}\partial ^{\mu }A_{\mu }$, which is a pseudo-differential
operator \cite{pdif}.

The quantum-particle of this field is called Podolsky photon and the
interaction between these particles and fermionic (scalar) fields is known
as (scalar) generalized quantum electrodynamics, GQED$_{4}$ (GSQED$_{4}$)
\cite{bufalo1, bufalo2, bufalo3}. It was shown in these series of papers
that the Podolsky photon has the quality of controlling UV divergencies,
leading to an almost finite theory, since the finiteness depends on the
divergence degree of the diagram. Further analysis with Bopp-Podolsky
electrodynamics were realized at thermodynamical equilibrium \cite{Boni1,
Boni2}, with boundary conditions \cite{Casimir}, and in the presence of
external sources \cite{FGA}. Furthermore, in previous analysis the authors
have determined a bound for the free parameter $m_{p}\geq 350$ GeV \cite%
{bufalo2, Daniel}.

In particular, it should be noted that higher-derivative theories have a
Hamiltonian which is not bounded from below \cite{ref7} and that the
addition of such terms leads to the existence of negative norm states (or
ghosts states) jeopardizing thus the unitarity \cite{ref8}. Despite the fact
that many attempts to restore the unitarity by means of overcoming these
ghost states, no one has been able to give a general method to deal with
them \cite{ref9,ref10}. Nonetheless, recently in Ref.~\cite{Kaparulin} a
procedure was suggested for including interactions in free HD systems
without breaking their stability, remarkably it was shown that the dynamics of the
GQED is stable at both classical and quantum level.

Therefore, based on the positive aspects and outcome of both Bopp-Podolsky
electrodynamics and DKP theory, it is rather natural a systematic study of
their interaction. Moreover, this work concerns itself in the scalar sector
of the DKP theory, describing the eletromagnetic interaction between scalar
fields in a different phenomenologically way, which will certainly
complement the known results in the literature \cite{kinoshita, bufalo3}.
This work is therefore devoted to the analysis of the Generalized Scalar
Duffin-Kemmer-Petiau Electrodynamics. In Sec.~\ref{sec1} the covariant
transition amplitude is derived by a constraint analysis in the non-mixing
Lorentz gauge condition. In Sec.~\ref{sec2} the Schwinger-Dyson-Fradkin
equations are calculated, and the complete expressions for the basic Green's
functions is obtained. In Sec.~\ref{sec3} the Ward-Takahashi-Fradkin
identities are derived and subsequently, in Sec.~\ref{sec4}, the
renormalizability of the present theory is established. In Sec.~\ref{sec5}
the radiative corrections at one loop are computed and the (finite)
counter-terms are presented. In Sec.~\ref{sec6} the photon propagator and
the vertex at two-loops order are discussed diagrammatically. In Sec.~\ref%
{sec7} the authors present their final remarks and prospects. In the whole
work the metric signature $(+,-,-,-)$ for the Minkowski spacetime is used.


\section{Canonical transition amplitude}

\label{sec1}

The Lagrangian density describing the GSDKP is defined by\footnote{%
Throughout the text the following compact notation $\hat{O}%
=\beta ^{\mu }O_{\mu }$ will be used.}
\begin{equation}
\mathcal{L}=\frac{i}{2}\bar{\psi}\beta ^{\mu }\left( \partial _{\mu }\psi
\right) -\frac{i}{2}\left( \partial _{\mu }\bar{\psi}\beta ^{\mu }\right)
\psi -m\bar{\psi}\psi +eA_{\mu }\bar{\psi}\beta ^{\mu }\psi -\frac{1}{4}%
F_{\mu \nu }F^{\mu \nu }+\frac{a^{2}}{2}\partial ^{\mu }F_{\mu \beta
}\partial _{\alpha }F^{\alpha \beta },  \label{eq6}
\end{equation}%
where $F_{\mu \nu }=\partial _{\mu }A_{\nu }-\partial _{\nu }A_{\mu }$ is
the usual electromagnetic field-strength tensor and $\beta ^{\mu }$ are the
DKP matrices that obey the algebra
\begin{equation}
\beta ^{\mu }\beta ^{\nu }\beta ^{\theta }+\beta ^{\theta }\beta ^{\nu
}\beta ^{\mu }=\beta ^{\mu }\eta ^{\nu \theta }+\beta ^{\theta }\eta ^{\nu
\mu }.  \label{dkpalgebra}
\end{equation}%
For further information about the representation of the matrices $\beta
^{\mu }$ see the appendix \ref{sec:B}. Although DKP theory is formally
similar to Dirac theory, there are several subtle contrasting behaviour
already in classical level. For instance, the conjugated field of the fermionic theory is characterized as $\bar{\psi}=\psi ^{\dag }\gamma ^{0}$, whereas the
conjugate DKP field is defined such as $\bar{\psi}=\psi ^{\dag }\eta ^{0}$,
where $\eta ^{0}=2(\beta ^{0})^{2}-1$. Moreover, one can sample for an
arbitrary four-vector $p$ the following relation is satisfied
\begin{equation}
\hat{p}(\hat{p}^{2}-p^{2})=0.
\end{equation}%
This shows another contrast with the fermionic theory, since $\hat{p}%
^{2}\neq p^{2}$. Nonetheless, this relation combined with plane wave
solutions for the free field equations leads to $p^{2}=m^{2}$.

Classically this theory is invariant under local gauge transformations%
\begin{equation}
\psi \rightarrow e^{i\alpha \left( x\right) }\psi ,\quad A_{\mu }\rightarrow
A_{\mu }+\frac{1}{e}\partial _{\mu }\alpha \left( x\right) .
\end{equation}
The Euler-Lagrange equations are obtained as usual from the Hamiltonian
principle
\begin{align}
\left[ i\beta ^{\mu }(\partial _{\mu }-ieA_{\mu })-m\right]\psi &=0 , \\
(1+a^{2}\square )\partial _{\mu }F^{\lambda \mu }&=e\bar{\psi}\beta
^{\lambda }\psi .
\end{align}
The translational space-time invariance of the Lagrangian density leads to the canonical Hamiltonian
\begin{align}
H_{c}=& \int d^{3}x \bigg[(\mathcal{\partial }_{0}\mathcal{\bar{\psi}})\frac{%
\mathcal{\partial L}}{\partial (\partial _{0}\mathcal{\bar{\psi})}}+\frac{%
\mathcal{\partial L}}{\partial (\partial _{0}\mathcal{\psi )}}(\mathcal{%
\partial }_{0}\mathcal{\psi })\mathcal{+}\frac{\mathcal{\partial L}}{%
\partial (\partial _{0}A_{\nu }\mathcal{)}}(\mathcal{\partial }_{0}A_{\nu })
\notag \\
&-\partial _{\theta }\left( \frac{\mathcal{\partial L}}{\partial (\partial
_{0}\partial _{\theta }A_{\nu }\mathcal{)}}\right) (\mathcal{\partial }%
_{0}A_{\nu }) +\frac{\mathcal{\partial L}}{\partial (\partial _{0}\partial
_{\theta }A_{\nu }\mathcal{)}}(\partial _{\theta }\mathcal{\partial }%
_{0}A_{\nu })- \mathcal{L} \bigg].
\end{align}
Thus the canonical momenta associated with the DKP fields $\left(\bar{\psi}%
,\psi\right)$ are
\begin{align}
p &=\frac{\mathcal{\partial L}}{\partial (\partial _{0}\mathcal{\bar{\psi})}}%
=-\frac{i}{2}\beta ^{0}\psi, \\
\bar{p}&=\frac{\mathcal{\partial L}}{\partial (\partial _{0}\mathcal{\psi )}}%
=\frac{i}{2}\bar{\psi}\beta ^{0} ,  \label{vinpri1}
\end{align}
whereas the canonical momenta for gauge fields are obtained from the
Ostrogradski method \cite{Ostro}. This method consists in defining the
dynamics of the system in a first-order form, i.e., the dynamics takes place
in a spanned phase space characterized by the independent variables $%
A_{\mu}, \Pi ^{\nu }$ and $\Gamma_{\mu} \equiv \partial_0 A_{\mu}, \Phi
^{\nu }$
\begin{align}
\Pi ^{\nu }&= \frac{\partial\mathcal{\ L}}{\partial \Gamma_{\nu} } -
2\partial _k \frac{\partial \mathcal{L}}{\partial \left(\partial _{k}
\Gamma_{\nu}\right) } -\partial _{0} \frac{\partial \mathcal{L}}{\partial
\left(\partial _{0} \Gamma_{\nu}\right) },  \notag \\
& = F^{\nu 0}+a^{2}[\eta ^{i\nu }\partial _{i}\partial _{\alpha }F^{\alpha
0}-\partial _{0}\partial _{\alpha }F^{\alpha \nu }] , \\
\Phi ^{\nu }&= \frac{\partial \mathcal{L}}{\partial \left(\partial _{0}
\Gamma_{\nu}\right) } =a^{2}[\partial _{\alpha }F^{\alpha \nu }-\eta ^{\nu
0}\partial _{\alpha }F^{\alpha 0}].  \label{vinpri2}
\end{align}
From the above momentum expressions, the constraint structure
of the theory can be studied by following Dirac's approach to singular systems \cite{dirac}.
In this approach it is possible to obtain the set of first-class constraints
\begin{align}
\varphi _{1}=\Phi _{0}\approx 0 , \quad \varphi _{2}= \Pi _{0}-\partial
_{k}\Phi ^{k}\approx 0 , \quad \varphi _{3}= e\bar{\psi}\beta
^{0}\psi-\partial ^{k}\Pi _{k}\approx 0 ,
\end{align}
and the set of second-class constraints
\begin{align}
\chi ^{(1)}&=p+\frac{i}{2}\beta ^{0}\psi \approx 0 , \quad \bar{\chi}^{(1)} =%
\bar{p}-\frac{i}{2}\bar{\psi}\beta ^{0}\approx 0 , \\
\chi ^{(2)}& =[1-(\beta ^{0})^{2}][i\beta ^{i}\partial _{i}\psi (x)-m\psi
(x)+e\beta ^{i}A_{i}(x)\psi (x)]\approx 0 , \\
\bar{\chi}^{(2)}&=[-i\partial _{i}\bar{\psi}(x)\beta ^{i}+m\bar{\psi}(x)-e%
\bar{\psi}(x)\beta ^{i}A_{i}(x)][1-(\beta ^{0})^{2}]\approx 0.
\end{align}
The weak equality $\approx $ is understood in according to Dirac's sense.

With the full set of first-class and second-class constraints determined, the next step is to obtain the functional generator. The transition
amplitude in the Hamiltonian form is written in the following way \cite{FSej}
\begin{equation}
Z=N\int D\mu \exp \left\{ i\int d^{4}x\left[ \left( \mathcal{\partial }_{0}%
\mathcal{\bar{\psi}}\right) p+\bar{p}\left( \mathcal{\partial }_{0}\mathcal{%
\psi }\right) \mathcal{+}\Pi ^{\nu }\left( \mathcal{\partial }_{0}A_{\nu
}\right) +\Phi ^{\nu }\left( \partial _{0}\Gamma _{\nu }\right) \mathcal{-H}%
_{c}\right] \right\}
\end{equation}%
where the canonical Hamiltonian is given by
\begin{align}
\mathcal{H}_{c}&=\Pi _{0}\Gamma ^{0}+\Pi _{k}\Gamma ^{k}+\Phi _{k}(\partial
^{k}\Gamma _{0}-\partial _{l}F^{lk}+\frac{\Phi ^{k}}{2a^{2}})-\frac{i}{2}%
\bar{\psi}\beta ^{i}\overleftrightarrow{\partial }_{i}\psi +m\bar{\psi}\psi
\notag \\
&-e\bar{\psi}\hat{A}\psi +\frac{1}{4}F_{kj}F^{kj}+\frac{1}{4}(\Gamma
_{j}-\partial _{j}A_{0})^{2}-\frac{a^{2}}{2}(\partial ^{j}\Gamma
_{j}-\partial ^{j}\partial _{j}A_{0})^{2},
\end{align}
and the integration measure is defined in such a way that it transforms as
an scalar at the constrained phase space%
\begin{equation}
D\mu = D\Phi ^{\nu }D\Gamma _{\nu }D\Pi ^{\mu }DA_{\mu }D\mathcal{\bar{\psi}}%
D\mathcal{\psi }D\bar{p}Dp\delta (\Theta _{l})\det \left\Vert \left\{\Theta
_{l},\Theta _{m} \right\}\right\Vert ^{\frac{1}{2}}.
\end{equation}%
Now, the complete set of constraints for the GSDKP is
\begin{equation}
\Theta _{l}=\left\{ \chi ^{(1)},\bar{\chi}^{(1)},\chi ^{(2)},\bar{\chi}%
^{(2)},\varphi _{1},\varphi _{2},\varphi _{3},\Sigma _{1},\Sigma _{2},\Sigma
_{3}\right\} ,
\end{equation}
in which a suitable gauge conditions for the first-class constraints are
chosen as the generalized radiation conditions \cite{galvao}
\begin{equation}
\Sigma _{1}=\Gamma _{0}(x)\approx 0, \quad \Sigma _{2}=A_{0}\approx 0, \quad
\Sigma _{3}=(1+a^{2}\square )(\vec{\nabla}.\vec{A})\approx 0.
\end{equation}

After integrating over the gauge and fermionic momenta the transition
amplitude Z is explicitly written
\begin{align}
Z&=N\int DA_{\mu }D\mathcal{\bar{\psi}}D\mathcal{\psi }\det \left\Vert
(1+a^{2}\vec{\nabla}^{2})\vec{\nabla}^{2}\right\Vert \delta ((1+a^{2}\square
)(\vec{\nabla}.\vec{A}))  \notag \\
& \times \exp \left[ i\int d^{4}x \left\{\mathcal{\bar{\psi}}(i\beta ^{\mu }\nabla _{\mu
}-m)\mathcal{\psi }-\frac{1}{4}F_{\mu \nu }F^{\mu \nu }+\frac{a^{2}}{2}%
\partial ^{\mu }F_{\mu \beta }\partial _{\alpha }F^{\alpha \beta }\right\} \right].
\label{amplicg}
\end{align}
Although the above expression is correct its form is not explicitly
covariant; then it is not convenient for purposes of calculation.
However the Faddeev-Popov-DeWitt ansatz \cite{faddeev} allows a covariant form for the amplitude of vacuum-vacuum transition.

Hence, using the Faddeev-Popov-DeWitt ansatz in the non-mixing gauge
condition \cite{LW}
\begin{equation}
\Omega \left( A\right) =\left( 1+a^{2}\square \right) ^{\frac{1}{2}}\partial
^{\mu }A_{\mu }
\end{equation}%
the transition amplitude can be written as
\begin{align}
Z&=N\int DA_{\mu }D\mathcal{\bar{\psi}}D\mathcal{\psi } \exp \bigg\{ i\int
d^{4}x \bigg[ \mathcal{\bar{\psi}}\left( i\beta ^{\mu }\nabla _{\mu }-m\right)
\mathcal{\psi }-\frac{1}{4}F_{\mu \nu }F^{\mu \nu }  \notag \\
&+\frac{a^{2}}{2}\partial ^{\mu }F_{\mu \beta }\partial _{\alpha }F^{\alpha
\beta }-\frac{1}{2\xi }\left( \partial ^{\mu }A_{\mu }\right) \left(
1+a^{2}\square \right) \left( \partial ^{\mu }A_{\mu }\right) \bigg] \bigg\}.
\end{align}
The choice of using the non-mixing gauge condition, $\left( 1+a^{2}\square
\right) ^{\frac{1}{2}}\partial ^{\mu }A_{\mu } $, is rather justified by
calculation purposes because it preserves the order of the field equation
\cite{Daniel}; since the natural choice in the Podolsky theory, the
generalized Lorenz term $(1+a^{2}\square )(\partial _{\mu }A^{\mu })$
complicates the theory's quantization once it increases the order of the
field equation. Then the non-mixing gauge term is related to a
pseudodifferencial operator \cite{pdif}.

The minimal coupling DKP functional generator with the higher-derivative
Podolsky term can be written as
\begin{equation}
\mathcal{Z}\left[ \eta ,\bar{\eta},J_{\mu }\right] =\int D\mu \left( \psi ,%
\bar{\psi},A_{\mu }\right) \exp \left[ iS_{eff}\right]  \label{func gerador}
\end{equation}%
where the effective action is defined by%
\begin{align}
S_{eff}&=\int d^{4}x \bigg[\bar{\psi}\left( i\beta ^{\mu }\partial _{\mu
}-m+e\beta ^{\mu }A_{\mu }\right) \psi -\frac{1}{4}F_{\mu \nu }F^{\mu \nu }+%
\frac{a^{2}}{2}\partial ^{\mu }F_{\mu \beta }\partial _{\alpha }F^{\alpha
\beta }  \notag \\
&-\frac{1}{2\xi } \left( \partial ^{\mu }A_{\mu }\right) \left(
1+a^{2}\square \right) \left( \partial ^{\mu }A_{\mu }\right)  +\bar{%
\psi}\eta +\bar{\eta}\psi +A^{\mu }J_{\mu }\bigg].  \label{acao}
\end{align}
and $\eta $, $\bar{\eta}$ and $J_{\mu }$ are the sources from fundamental
fields involved, namely $A_{\mu }$, $\psi $ e $\bar{\psi}$.


\section{Schwinger-Dyson-Fradkin equations}

\label{sec2}

It has been known for a long time that it is possible to describe all
content of a particular field theory as a set of field equations in the
Heisenberg description. The most elegant way of studying such equations and
extract the physical content is the functional formulation, consisting in
an infinite chain of differential equations that relates different Green's
function in an exact manner \cite{nash, Fradkin}. This infinite tower of
equations refers to the Schwinger-Dyson-Fradkin (SDF) equations.

The propose of this section is to determine the complete SDF equations for the basic
propagators, for the gauge and DKP fields, and also for the vertex function using the functional generator defined by the equation %
\eqref{func gerador}.

\subsection{The Schwinger-Dyson-Fradkin equations for the photon propagator}

The complete expression for the gauge-field propagator can be determined by
means of the functional generator \eqref{func gerador} leading to the
Schwinger variational equation for the gauge field, in which $\mathrm{S}$
differs from $\mathrm{S}_{eff}$ by source terms
\begin{equation}
\left[ \left. \frac{\delta \mathrm{S}}{\delta A_{\gamma }\left( x\right) }%
\right\vert _{ \frac{\delta}{\delta i\eta } ,\frac{\delta}{\delta i\bar{\eta}
} ,\frac{\delta}{\delta iJ_{\mu } }} +J^{\gamma }\left( x\right) \right]
\mathcal{Z}\left[ \eta ,\bar{\eta},J_{\mu }\right] =0 .
\end{equation}
It should be remarked that the field limits are related to the functional
Fourier transform as
\begin{equation}
A_{\gamma }\left( x\right) \rightarrow \frac{1}{i}\frac{\delta }{\delta
J^{\gamma }\left( x\right) },\quad \bar{\psi}\left( x\right) \rightarrow
\frac{1}{i}\frac{\delta }{\delta \eta \left( x\right) },\quad \psi \left(
x\right) \rightarrow \frac{1}{i}\frac{\delta }{\delta \bar{\eta}\left(
x\right) } .  \label{aderivadaj}
\end{equation}%
Nonetheless, in solving the above relation the generating
functional $\mathcal{Z}\left[ J\right] =\exp \left\{ iW\left[ J\right]
\right\} $ for the connected Green's functions must be introduced. Then the
Schwinger variational equation becomes
\begin{align}
-J^{\gamma }\left( x\right) =&-ie\frac{\delta }{\delta \eta \left( x\right) }%
\beta ^{\gamma }\left( \frac{\delta W}{\delta \bar{\eta}\left( x\right) }%
\right) +e\frac{\delta W}{\delta \eta \left( x\right) }\beta ^{\gamma }\frac{%
\delta W}{\delta \bar{\eta}\left( x\right) }  \notag \\
& +\left[ T^{\gamma \mu }+\frac{1}{\xi }L^{\gamma \mu }\right] \left(
1+a^{2}\square \right) \square \frac{\delta W}{\delta J^{\mu }\left(
x\right) }.  \label{func green conectadas}
\end{align}
The last equation can be interpreted as the complete Podolsky field equation
subjected to a external source $J^{\gamma }$. On this equation $T^{\gamma
\mu }$ and $L^{\gamma \mu }$ are differential projectors
\begin{equation}
T^{\gamma \mu }+L^{\gamma \mu }=g^{\gamma \mu }, \quad L^{\gamma \mu }=\frac{%
\partial ^{\gamma }\partial ^{\mu }}{\square }.
\end{equation}

In order to obtain the complete gauge-field propagator it proves convenient
to introduce the generating functional for the one particle irreducible
(1PI) Green's functions as well, which is related to $W$ by a functional Legendre
transformation
\begin{equation}
\Gamma \left[ \psi ,\bar{\psi},A_{\mu }\right] =W\left[ \eta ,\bar{\eta}%
,J_{\mu }\right] -\int d^{4}x\left( \bar{\psi}\eta +\bar{\eta}\psi +A^{\mu
}J_{\mu }\right).  \label{1PI}
\end{equation}%
Hence, rewriting \eqref{func green conectadas} in terms of the 1PI $\Gamma %
\left[ \psi ,\bar{\psi},A_{\mu }\right] $ and differentiating the resulting
expression with respect to $A_{\nu }\left( y\right) $%
\begin{align}
\frac{\delta ^{2}\Gamma }{\delta A_{\nu }\left( y\right) \delta A_{\gamma
}\left( x\right) }&=-ie\beta ^{\gamma }\frac{\delta }{\delta A_{\nu }\left(
y\right) }\left( \frac{\delta ^{2}W}{\delta \eta \left( x\right) \delta \bar{%
\eta}\left( x\right) }\right) +\left[ T^{\gamma \nu }+\frac{1}{\xi }%
L^{\gamma \nu }\right] \left( 1+a^{2}\square \right) \square \delta ^{\left(
4\right) }\left( x,y\right) .  \label{irredutivel}
\end{align}
From the above definitions, one can obtain identities relating the connected
and 1PI two-point functions. For instance, it follows that for the DKP field
\begin{equation}
i\int d^{4}z\mathcal{S}\left( x,z;A\right) \frac{\delta ^{2}\Gamma }{\delta
\psi \left( y\right) \delta \bar{\psi}\left( z\right) }=\delta ^{\left(
4\right) }\left( x-y\right) ,  \label{identidade}
\end{equation}%
in which the complete DKP propagator is defined such as
\begin{equation}
\mathcal{S}\left( x,z;A\right)= i\left. \frac{\delta ^{2}W\left[ \eta ,\bar{%
\eta},J_{\mu }\right] }{\delta \eta \left( z\right) \delta \bar{\eta}\left(
x\right) }\right\vert _{\eta =\bar{\eta}=0} .  \label{funcional}
\end{equation}%
Another important quantity to be defined is the complete DKP-photon 1PI
vertex function%
\begin{equation}
e\Gamma ^{\nu }\left( x,z;y\right) = \left. \frac{\delta ^{3}\Gamma }{\delta
A_{\nu }\left( y\right) \delta \psi \left( z\right) \delta \bar{\psi}\left(
x\right) }\right\vert _{A_{\nu }=\psi =\bar{\psi}=0},  \label{1pifdkp}
\end{equation}%
which after some algebraic manipulation makes the equation %
\eqref{irredutivel} possible to be rewritten as%
\begin{align}
\frac{\delta ^{2}\Gamma }{\delta A_{\nu }\left( y\right) \delta A_{\gamma
}\left( x\right) }=&\left[ T^{\gamma \nu }+\frac{1}{\xi }L^{\gamma \nu }%
\right] \left( 1+a^{2}\square \right) \square \delta ^{\left( 4\right)
}\left( x,y\right)  \notag \\
&+ie^{2}\int d^{4}ud^{4}wTr\left[ \mathcal{S}\left( x,u;A\right) \beta
^{\gamma }\mathcal{S}\left( w,x;A\right) \Gamma ^{\nu }\left( u,w;y\right) %
\right] .  \label{tchegando}
\end{align}
The second term of \eqref{tchegando} can be identified with the polarization
operator, $\Pi ^{\gamma \nu }$,
\begin{equation}
\Pi ^{\gamma \nu }\left( x,y\right) =ie^{2}\int d^{4}ud^{4}wTr\left[
\mathcal{S}\left( x,u;A\right) \beta ^{\gamma }\mathcal{S}\left(
w,x;A\right) \Gamma ^{\nu }\left( u,w;y\right) \right],
\label{autoenergia foton}
\end{equation}%
defined as the sum of all compact self-energy photon parts. The absence of a
$\left( -1\right) $ factor comes from the fact that there is a bosonic loop
related to the DKP field, not a fermionic as in the Dirac field.

\begin{figure}
\centering
\includegraphics[width=15cm]{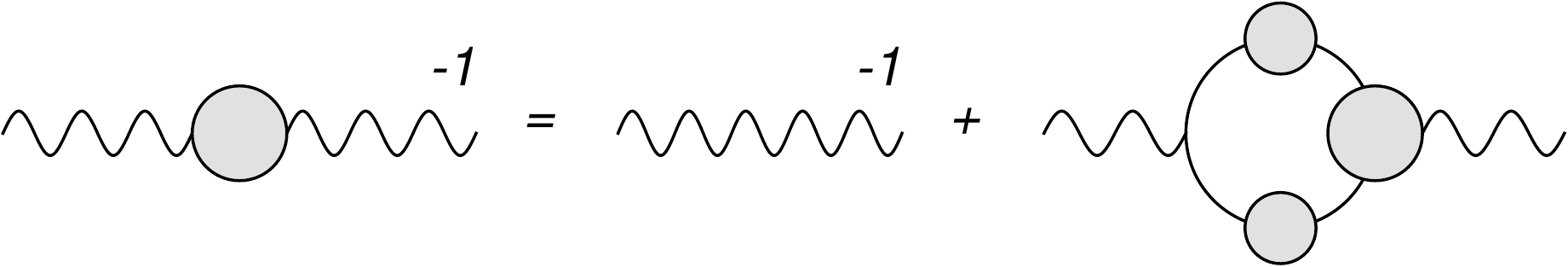}
\caption{The SDF equation for the photon propagator.}
\label{fig1}
\end{figure}

Then the gauge field satisfies an identity as \eqref{identidade}; therefore
\begin{equation}
(\mathfrak{D}^{\nu \rho}) ^{-1} \left( z,y\right)= \frac{\delta ^{2}\Gamma }{%
\delta A_{\rho }\left( y\right) \delta A_{\nu }\left( z\right) },
\label{iddentidade}
\end{equation}%
relates the inverse of the complete (and free) photon propagator to the 1PI
Green's function. Thus the expression for the photon's inverse
complete propagator in momentum representation is
\begin{equation}
(\mathfrak{D}^{\gamma \nu })^{-1} \left( p\right) =(D^{\gamma \nu
})^{-1}\left( p\right) +\Pi ^{\gamma \nu }\left( p\right).
\label{inversa do full propag foton}
\end{equation}%
We can represent the above equation diagrammatically, as in Figure \ref{fig1}.

The expression \eqref{inversa do full propag foton} can be solved in order
to find
\begin{equation}
i\mathfrak{D}^{\gamma \nu }\left( p\right) =-\frac{\left( \eta ^{\gamma \nu
}-\frac{p^{\gamma }p^{\nu }}{p^{2}}\right) }{\left[ \Pi \left( p\right)
-\left( 1-a^{2}p^{2}\right) p^{2}\right] }+\frac{\xi }{p^{2}\left(
1-a^{2}p^{2}\right) }\frac{p^{\gamma }p^{\nu }}{p^{2}},
\label{full prop foton}
\end{equation}%
in which the scalar polarization $\Pi \left( p\right) $ is related to the
scalar polarization $\Pi ^{\gamma \nu }\left( p\right) $ through the
structure
\begin{equation}
\Pi ^{\gamma \nu }\left( p\right) =\left( -p^{2}\eta ^{\gamma \nu
}+p^{\gamma }p^{\nu }\right) \Pi \left( p\right) .
\end{equation}

For the free propagator, namely $\Pi \left( p\right) =0$ on %
\eqref{full prop foton} and $a=m_{p}^{-1}$, one has the expression%
\begin{equation}
iD^{\gamma \nu }\left( p\right) =\left[ \eta ^{\gamma \nu }-\left( 1-\xi
\right) \frac{p^{\gamma }p^{\nu }}{m_{p}^{2}}\right] \left[ \frac{1}{p^{2}}-%
\frac{1}{p^{2}-m_{p}^{2}}\right] -\left( 1-\xi \right) \frac{p^{\gamma
}p^{\nu }}{\left( p^{2}\right) ^{2}}.  \label{final}
\end{equation}%
Note that there are no mixing between the massless and massive poles, in contrast with the usual generalized Lorenz condition, owing to the non-mixing gauge fixing. It should be remarked
that in Ref.~\cite{Kaparulin} a procedure was suggested for including
interactions in free HD systems without breaking their stability (ghosts
modes) and it holds for GSDKP. In addition, previous results
in the fermionic and mesonic generalized theories \cite%
{bufalo1,bufalo2,bufalo3} also motivate an attention to the present
theory, once the propagator \eqref{final} has a UV finite behavior (in the light of effective
theories) and an interesting renormalized behavior.


\subsection{The Schwinger-Dyson-Fradkin equations for the DKP propagator}

\begin{figure}
\centering
\includegraphics[width=15cm]{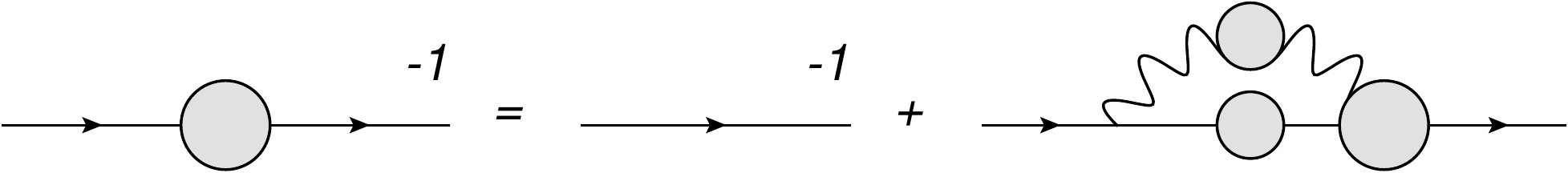}
\caption{The SDF equation for the scalar propagator.}
\label{fig2}
\end{figure}

This subsection is devoted to keep on deriving the SDF equations,
obtaining now an integral expression for the complete DKP propagator. Starting with the Schwinger
variational equation%
\begin{equation}
\left[ \left. \frac{\delta S}{\delta \bar{\psi}\left( x\right) }\right\vert
_{ \frac{\delta}{\delta i\eta } ,\frac{\delta}{\delta i\bar{\eta} } ,\frac{%
\delta}{\delta iJ_{\mu } }} +\eta \left( x\right) \right] \mathcal{Z}\left[
\eta ,\bar{\eta},J_{\mu }\right] =0 .  \label{var schw dkp}
\end{equation}%
writing it in terms of the generating
functional $W$ and then differentiating the resulting expression with
respect to the source $\eta \left( y\right) $ leads to
\begin{equation}
i\delta ^{\left( 4\right) }\left( x-y\right) =-\left[ i\beta ^{\mu }\partial
_{\mu }-m+e\beta ^{\mu }\left\langle A_{\mu }\right\rangle -ie\beta ^{\mu }%
\frac{\delta }{\delta J^{\mu }\left( x\right) }\right] \mathcal{S}%
\left(x,y;A\right) .  \label{eq4}
\end{equation}%
by solving the derivative of the last term one can immediately identify
\begin{equation}
\Sigma \left( x,z\right) =ie^{2}\beta ^{\mu }\int d^{4}ud^{4}v\mathfrak{D}%
_{\mu }^{\alpha }\left( u,x\right) \mathcal{S}\left( x,v;A\right) \Gamma
_{\alpha }\left( v,z;u\right) ,  \label{sigma}
\end{equation}%
as the DKP self-energy function $\Sigma $. Hence, by taking the limit of
null sources
\begin{equation}
i\delta ^{\left( 4\right) }\left( x-y\right) =-\left[ i\beta ^{\mu }\partial
_{\mu }-m\right] \mathcal{S}\left( x,y;A\right) +\int d^{4}z\Sigma \left(
x,z\right) \mathcal{S}\left( z,y;A\right) .
\end{equation}%
In momentum representation this equation becomes $\mathcal{S}^{-1}\left(
p\right) =S^{-1}(p)+\Sigma \left( p\right) $ . This equation can be viewed
as in the figure \ref{fig2}.

The above equation can formally be written as
\begin{equation}
\mathcal{S}\left( p\right) =\frac{i}{\beta ^{\mu }p_{\mu }-\mathfrak{M}
\left( p\right) }  \label{propag}
\end{equation}%
where the mass operator $\mathfrak{M}$ is defined by
\begin{equation}
\mathfrak{M}(p)=m+\Sigma (p)  \label{mass operator}
\end{equation}%
showing that the mass operator encompasses both the DKP self-energy $\Sigma $
and the bare mass $m$.

Besides, the expression for the DKP free propagator can be obtained with the
help of the DKP algebra \eqref{dkpalgebra},
\begin{equation}
S(p)=i\frac{1}{m}\left[\frac{\hat{p}(\hat{p}+m)}{(p^{2}-m^{2})}-1\right].
\end{equation}

As one can see, the self-energy function \eqref{sigma}, differently from the
photon function \eqref{autoenergia foton}, is sensitive to the effects of
the Podolsky $m_P$-dependent terms of \eqref{final} already at first order
on perturbation theory.


\subsection{The Schwinger-Dyson-Fradkin equations for the vertex part}

As is well known, the SDF equations do not only depend on the fundamental
Green's functions of a given theory, but they do depend on higher-order
functionals, which also satisfy their own SDF equations. This will become
clear in the derivation of the vertex function. Although it should be
remarked that it is possible to find a relation that
connects the complete vertex function with $\mathcal{S}$ and $\mathcal{D}$
which contain only skeleton graphs, i.e., connected graphs \cite{Landau, EQC}.

The starting point for the derivation of the vertex function is \eqref{eq4}, this also follows from the guideline presented previously. In a
similar way, on taking the derivative of the resulting expression with respect
to the field $A_{\sigma}(z)$, and after some manipulations, one finds the
following expression for the vertex function
\begin{equation}
i\Gamma ^{\sigma }\left( q,p;k\right) =-\beta ^{\sigma }(2\pi )^{4}\delta
\left( q-p-k\right) +i\Lambda ^{\sigma }\left( k,p;q\right)  \label{gamis}
\end{equation}%
in which a new quantity, the vertex part, has been introduced
\begin{align}
i\Lambda ^{\sigma }\left( q,p;k\right) &=ie^{2}\beta ^{\mu }\frac{1}{\left(
2\pi \right) ^{4}}\int d^{4}t\mathfrak{D}_{\mu \rho }\left( t\right)
\mathcal{S}\left( t+k\right) \Phi ^{\sigma \rho }\left( t+k,p;q,t\right)
\notag \\
&+e^{2}\beta ^{\mu }\frac{1}{\left( 2\pi \right) ^{4}}\int
d^{4}p_{1}d^{4}p_{2}\mathfrak{D}_{\mu \rho }\left( p_{1}\right) \mathcal{S}%
\left( p_{1}+k\right) \Gamma ^{\sigma }\left( \left( p_{1}+k\right)
,p_{2};q\right) \mathcal{S}\left( p_{2}\right) \Gamma ^{\rho }\left(
p_{2},p;p_{1}\right),  \label{the end}
\end{align}
and defined the four-point vertex function
\begin{equation}
e^{2}\Phi ^{\sigma \rho }\left( a,w;z,s\right) =\frac{\delta ^{4}\Gamma }{%
\delta A_{\rho }\left( s\right) \delta A_{\sigma }\left( z\right) \delta
\psi \left( w\right) \delta \bar{\psi}\left( a\right) }.  \label{func 4 ptos}
\end{equation}

\begin{figure}
\centering
\includegraphics[width=15cm]{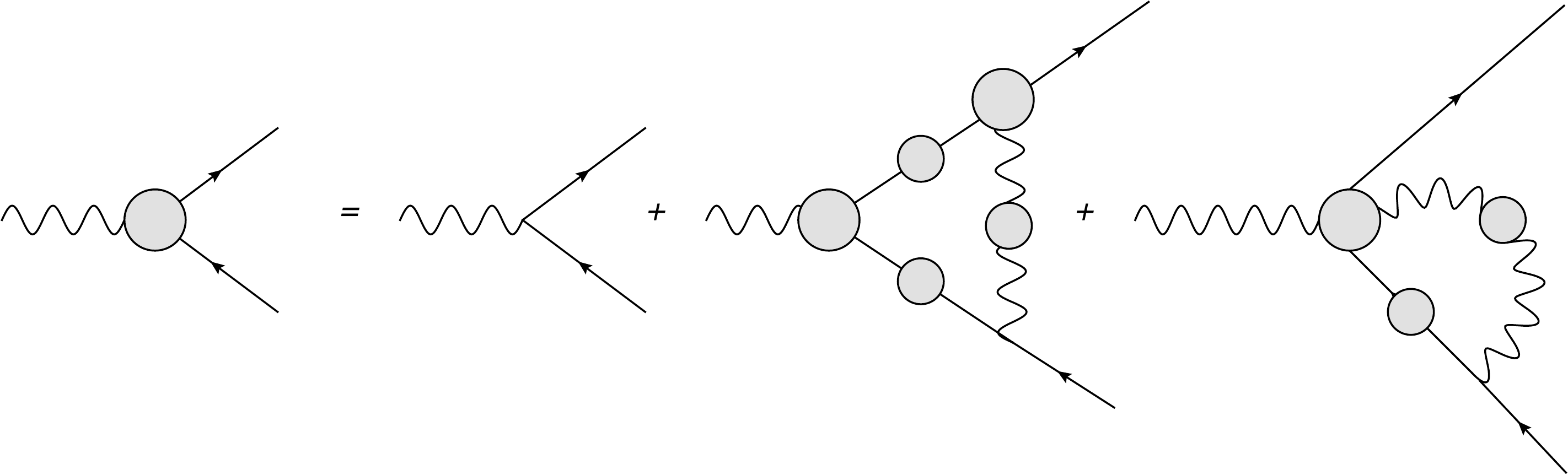}
\caption{The SDF equation for the vertex function.}
\label{fig3}
\end{figure}

Equation\eqref{the end} shows explicitly that the three-point vertex function
depends on the four-point one, emphasizing the tower of equations that SDF
equations are. However, the present work focuses in a perturbative
calculation, and then the situation here is not that complex, once the
three fundamental Green's functions of interest that can evaluate the
respective radiative corrections and the effects from the HD
contributions from the photon propagator had already been determined. Diagrammatically, the irreducible vertex part can be visualized in figure \ref{fig3}.

\vspace{1cm}


\section{Ward-Fradkin-Takahashi identities}

\label{sec3}

Although relativistic quantum systems are formulated in the framework of
gauge fields, all physical observables in a field theory are gauge
independent. The existence of a local gauge symmetry in a field theory
generates constraint relations between the theory's Green's functions. These
relations are known as the Ward--Fradkin--Takahashi identities. These
identities, in terms of Green's functions, are closely related with the
renormalizability of a theory. The purpose of this section is to derive such
identities for GSDKP electrodynamics using a functional approach \cite%
{pokorski}.

The derivation of the WTF identities is formally given in terms of the
following identity upon the functional generator \eqref{func gerador}
\begin{equation}
\left. \frac{\delta \mathcal{Z}\left[ \eta ,\bar{\eta},J_{\mu }\right]}{%
\delta \alpha (x)} \right| _{\alpha =0}=0.
\end{equation}
This leads to the equation of motion satisfied by $\mathcal{Z}\left[ \eta ,%
\bar{\eta},J_{\mu }\right]$
\begin{equation}
\left[ -i\frac{\square }{e\xi }\left( 1+a^{2}\square \right) \partial ^{\mu }%
\frac{\delta }{\delta J^{\mu }}-\frac{\delta }{\delta \eta }\eta +\bar{\eta}%
\frac{\delta }{\delta \bar{\eta}}-\frac{1}{e}\partial ^{\mu }J_{\mu }\right]
\mathcal{Z}=0.  \label{eq5}
\end{equation}%
Finally, one can obtain the desired quantum equation of motion for the
theory by writing \eqref{eq5} first in terms of $W$, and then as an
expression for the 1PI-generating functional $\Gamma \left[ \psi ,\bar{\psi}%
,A_{\mu }\right]$ through the relation \eqref{1PI}. One then obtain
\begin{equation}
-i\frac{\square }{e\xi }\left( 1+a^{2}\square \right) \partial _{x}^{\mu
}A_{\mu }-\bar{\psi}\frac{\delta \Gamma }{\delta \bar{\psi}}+\frac{\delta
\Gamma }{\delta \psi }\psi +\frac{1}{e}\partial _{x}^{\mu }\frac{\delta
\Gamma }{\delta A^{\mu }}=0.  \label{partida}
\end{equation}%
This is the equation that will supply all the WFT identities.

The first identity comes by applying the derivatives of \eqref{partida} with
respect to $\psi \left( y\right) $ and $\bar{\psi}\left( z\right) $,
yielding
\begin{equation}
\partial ^{\mu }\Gamma _{\mu }(z,y;x)=-\delta (x-z)\Gamma (x,y)+\Gamma
(x,z)\delta (x-y),  \label{wtfid}
\end{equation}%
where $\Gamma (x,z)= \frac{\delta ^2 \Gamma }{\delta \psi (z)\delta \bar{\psi%
} (x)}$. Besides, writing it in momentum representation
\begin{equation}
k_{\mu }\Gamma ^{\mu }\left( p,p^{\prime },k=p-p^{\prime }\right) =\mathcal{S%
}^{-1}\left( p^{\prime }\right) -\mathcal{S}^{-1}\left( p\right) .
\end{equation}%
Furthermore, on considering the limit of this equation as $k\rightarrow 0$,
one can find that the vertex part is related to the DKP self-energy
function as
\begin{equation}
\Lambda ^{\mu }\left( p,p,k=0\right) =-\frac{\partial }{\partial p_{\mu }}%
\Sigma \left( p\right) .  \label{eq1}
\end{equation}

On the other hand, upon the differentiation of \eqref{partida} with respect
to $A_{\nu}(y)$, it follows the identity
\begin{equation}
\partial _{\mu }\Gamma ^{\mu \nu }\left( x,y\right) =\frac{\square }{\xi }%
\left( 1+a^{2}\square \right) \partial ^{\nu }\delta ^{\left( 4\right)
}\left( x-y\right)
\end{equation}%
which, together with equation \eqref{autoenergia foton} implies that
\begin{equation}
k_{\mu }\Pi ^{\mu \nu }\left( k\right) =0 .  \label{transv}
\end{equation}%
Then, the longitudinal part does not take part of the dynamics in the sense that
it is not modified by radiative corrections.

The following section will present how the renormalization program
is implemented in the GSDKP, showing that by a renormalization of the fields
and physical quantities, such that the resultant, renormalized $S$--matrix
leads to finite values for all the processes.


\section{Renormalizability}

\label{sec4}

This section will include the on-shell renormalization program \cite%
{pokorski} for GSDKP electrodynamics. The following analysis will result in
state suitable physical conditions on the Green's functions serving as for
renormalization conditions, which shall be important to determinate the
renormalization constants (counterterms) in terms of (in)finite integrals as
well. Besides, the resulting renormalization condition on the DKP sector
will be more involving and subtle than the usual as in the Dirac theory,
because $\hat{p}^{2}\neq p^{2}$ in the DKP theory.

The bare Lagrangian density is defined in \eqref{eq6}. The standard
renormalization procedure begins introducing the renormalization constants
through the following replacements
\begin{equation}
\psi \rightarrow Z_{0}^{\frac{1}{2}}\psi , \quad A\rightarrow Z_{3}^{\frac{1%
}{2}}A .
\end{equation}
In this case the fully renormalized Lagrangian can be written as%
\begin{align}
\mathcal{L}&=\bar{\psi}(i\hat{\partial}-m+e\hat{A})\psi -\frac{1}{4}F_{\mu
\nu }F^{\mu \nu }+\frac{1}{2m_{p}^{2}}\partial ^{\mu }F_{\mu \beta }\partial
_{\alpha }F^{\alpha \beta }  \notag \\
&+\delta _{Z_{0}}\bar{\psi}i\hat{\partial}\psi -\delta _{Z_{1}}m\bar{\psi}%
\psi +\delta _{Z_{2}}e\bar{\psi}\hat{A}\psi -\frac{\delta _{Z_{3}}}{4}F_{\mu
\nu }F^{\mu \nu }  \label{Laget}
\end{align}
where the counterterms defined by $\delta _{Z_{i}}=Z_{i}-1$ were added, and the renormalization for the mass: $Z_{1} m= Z_{0}m_0$, and
for the vertex: $Z_{2}e =Z_{0}Z_{3}^{\frac{1}{2}} e_0 $ were also introduced. \footnote{%
The replacement $\bar{m}_{p}^2 =Z_3 m_{p}^2$ is only a matter of notation,
since there is not a renormalization constant associated with this parameter.%
}

From the renormalized Lagrangian \eqref{Laget} one can get the general
renormalized expressions for the SDF equations and WTF identities. In this scenario, on deriving the SDF equations one can conclude that the complete
propagators are changed as
\begin{equation}
\mathcal{D}^{\mu \nu }\rightarrow Z_{3}\mathcal{D}^{\mu \nu }, \quad
\mathcal{S}\rightarrow Z_{0}\mathcal{S}, \quad \Gamma ^{\mu }\rightarrow
Z_{2}^{-1}\Gamma ^{\mu }.
\end{equation}
Besides, from the WFT identity \eqref{wtfid} it follows the equality $%
Z_{0}=Z_{2}$, which are identically satisfied at all orders in perturbation
theory. This implies that the charge renormalization is determined only by $%
e =Z_{3}^{\frac{1}{2}} e_0 $.

The effects from the renormalization as in \eqref{Laget} into the radiative
corrections are that the self-energy functions previously derived are now
added by the counterterms $\delta _{Z_{i}}$. These new
self-energy functions are now denoted by the index $(R)$. Analysing first the
photon sector, for which the renormalized self-energy function reads
\begin{equation}
\Pi^{(R)}(p) =\Pi(p)+\delta _{Z_{3}},
\end{equation}
then $\Pi(p)$ is the polarization scalar written in terms of the
renormalized quantities.

The first renormalization condition imposes that the complete photon
propagator \eqref{full prop foton}, with $\xi =1$, behaves as a massless
field
\begin{equation}
iD^{\gamma \nu }\left( p\right) =\eta ^{\gamma \nu } \frac{1}{p^{2}}, \quad
\text{when} ~~p^2 \rightarrow 0.
\end{equation}%
By means of the above condition, the counterterm $\delta
_{Z_{3}}$ is determined by
\begin{equation}
\delta _{Z_{3}}=-\left. \Pi(p)\right| _{p^{2}\rightarrow 0}.
\label{fotoncontrater}
\end{equation}

The renormalization conditions in the DKP sector are easily imposed into the
two-point 1PI function $\Gamma ^{(R)}(p)=\hat{p}-m-\Sigma ^{(R)}(p)$. The
first on-shell condition is that the physical mass is a pole \footnote{$%
m_{f} $ is defined as the zero of the DKP two-point 1PI function.}
\begin{equation}
\Gamma ^{(R)}(p)=\hat{p}-m_{f}, \quad \text{when}~~\hat{p}\rightarrow m_{f},
\label{cond1}
\end{equation}%
where
\begin{equation}
\Sigma ^{(R)}(p)=\Sigma (p)-m\delta _{Z_{1}}I+\delta _{Z_{0}}\hat{p}.
\end{equation}%
In contrast with the fermionic theory, the condition $%
\left. \frac{\partial \Gamma ^{(R)}(p)}{\partial \hat{p}}\right\vert _{\hat{p%
}\rightarrow m_{f}}=1$ will not be taken, once the
trilinear DKP algebra \eqref{dkpalgebra} leads to $\hat{p}^{2}\neq
p^{2} $, but instead $\hat{p}^{3}=p^{2}\hat{p}$, which complicates substantially the
derivation in terms of $\hat{p}$ and $p^2$. Nonetheless, a convenient choice
for the second renormalization condition is given by
\begin{equation}
\beta _\mu \frac{\partial \Gamma ^{(R)}(p)}{\partial p_\mu} =\beta _\mu
\beta ^\mu, \quad \text{when}~~\hat{p}\rightarrow m_{f},  \label{cond2}
\end{equation}%
since $\beta _\mu \beta ^\mu$ has a scalar structure. These renormalization
conditions, Eqs.\eqref{cond1} and \eqref{cond2}, when multiplied by the l.h.s by $\beta _\nu$ and
the r.h.s. by $\beta^\nu$, imply into the following
expressions for the counterterm $\delta _{Z_{0}}$ \footnote{%
The following decomposition $\Sigma (p)= \hat{p} \Sigma _{2}(p^{2})+ I \Sigma _{1}(p^{2})$ \cite{fai} and
identity $\beta _\nu \beta _\mu \beta ^\mu \beta ^\nu =4 I$ were used.}
\begin{equation}
- \delta _{Z_{0}}=  \left. \Sigma _{2}(p^{2})\right\vert _{p^{2}\rightarrow
m_{f}^{2}} + \frac{m_{f}^{2}}{2}  \beta _\nu \left. \frac{\partial
\Sigma _{2}(p^{2})}{\partial p^{2}} \beta ^\nu\right\vert _{p^{2}\rightarrow
m_{f}^{2}}+ \frac{m_{f}^{2}}{2} \left.\beta _\nu \frac{\partial
\Sigma _{1}(p^{2})} {\partial p^{2}} \beta ^\nu \right\vert _{p^{2}\rightarrow
m_{f}^{2}},  \label{1cond}
\end{equation}
and for the counterterm $\delta _{Z_{1}}$
\begin{equation}
m  \delta _{Z_{1}}=  \left. \Sigma _{1}(p^{2})\right\vert _{p^{2} \rightarrow
m_{f}^{2}} - \frac{m_{f}^{3}}{2} \left. \beta _\nu  \frac{\partial
\Sigma _{2}(p^{2})}{\partial p^{2}} \beta ^\nu \right\vert _{p^{2}\rightarrow
m_{f}^{2}}  - \frac{m_{f}^{2}}{2}  \left. \beta _\nu \frac{\partial
\Sigma _{1}(p^{2})} {\partial p^{2}} \beta ^\nu \right\vert _{p^{2}\rightarrow
m_{f}^{2}}.  \label{2cond}
\end{equation}
Therefore, from Eqs.\eqref{1cond} and \eqref{2cond}, the related DKP sector renormalization constants $Z_{0}$ and $Z_{1}$,
respectively, can be computed in all orders of perturbation theory.

At last, in order to uncover the renormalization constants, notice that the constant $Z_{0}$ can be determined by considering
that the renormalized vertex function \eqref{gamis}, by the on-shell
condition: $p^2 = q^2 =m^2$ and at a null transferred momentum limit $%
k^2=(p-q)^2 \rightarrow 0$, is
\begin{equation}
\bar{u} \left(q\right) i\Gamma ^{\sigma }\left( q,p;0\right) u
\left(p\right)=-(2\pi )^{4} \beta ^{\sigma },  \label{gamis2}
\end{equation}
or, equivalently, the vertex part \eqref{the end} is such that
\begin{equation}
\bar{u} \left(q\right) i\Lambda ^{\sigma }\left( q,p;0\right)u
\left(p\right)=0.  \label{rever}
\end{equation}

With this section the formal development of the theory has been concluded.
Henceforth, the explicit evaluation of the radiative
correction expressions for the photon polarization tensor, DKP self-energy
and vertex part will be proceed. The main interest is in observing the effects from the
HD terms into the UV behavior of these DKP radiative corrections. For this
purpose, a detailed discussion on the divergent structure of each
contributions will be done by computing their respective counterterms.


\section{Radiative corrections at one loop}

\label{sec5}

Once established the renormalizability of the GSDKP electrodynamics and with
the Schwinger-Dyson-Fradkin equations for the main complete Green's
functions, it is time to determine the radiative corrections at the lowest
order in perturbation theory. The divergences that appear in radiative
corrections will be regularized by the dimensional regularization
proceeding, which preserves all symmetries of the theory, in particular the
gauge symmetry \cite{pokorski,Framp}.


\subsection{The photon self-energy}

\begin{figure}
\centering
\includegraphics[width=6cm]{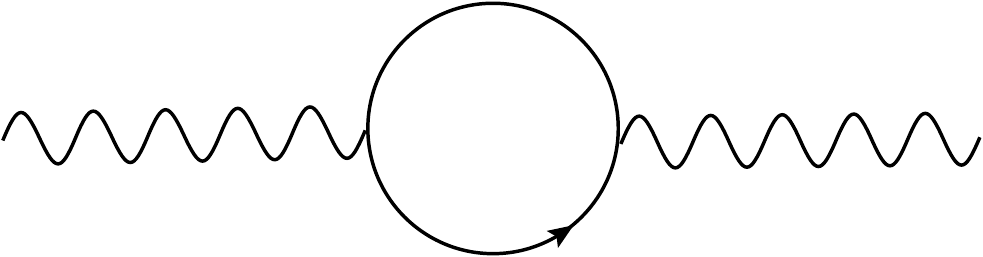}
\caption{Photon polarization tensor.}
\label{fig4}
\end{figure}

Let's start the study of radiative corrections for self-energy of the
photon. This quantity corresponds to the diagram shown in figure \ref{fig4}.

From the expression \eqref{autoenergia foton} rewritten in momentum
representation%
\begin{equation}
\Pi ^{\gamma \nu }\left( p\right) =ie^{2}\mu ^{4-d}\int \frac{d^{d}k}{(2\pi
)^{d}}Tr\left[ \beta ^{\gamma }S(p-k)\beta ^{\nu }S(-k)\right]
\end{equation}%
then
\begin{align}
\Pi ^{\gamma \nu }\left( p\right) =-\frac{ie^{2}\mu ^{4-d}}{m^2}\int \frac{%
d^{d}k}{(2\pi )^{d}}Tr\left\{ \beta ^{\gamma }\left[ \frac{\left( \hat{p}-%
\hat{k}\right) \left( \hat{p}-\hat{k}+m\right) }{\left( p-k\right) ^{2}-m^{2}%
}-1\right] \beta ^{\nu }\left[ \frac{-\hat{k}\left( -\hat{k} +m\right) }{%
k^{2}-m^{2}}-1\right] \right\} .  \label{aef}
\end{align}
Using the $\beta $ matrices trace properties \eqref{B3} the equation %
\eqref{aef} can be rewritten in the following form
\begin{align}
\Pi ^{\gamma \nu }\left( p\right) =\frac{-ie^{2}\mu ^{4-d}}{m^{2}}\int \frac{%
d^{d}k}{(2\pi )^{d}}\left\{ \frac{m^{2}\left( p-2k\right) ^{\gamma
}\left(p-2k\right) ^{\nu }-m^{2}\left[ \left( p-k\right) ^{2}+k^{2}-2m^{2}%
\right] \eta ^{\gamma \nu }}{\left[ \left( p-k\right) ^{2}-m^{2}\right]
\left( k^{2}-m^{2}\right) }\right\} .  \label{aefeq}
\end{align}
The momentum integration of the above terms can be performed by following
the well-known set of rules of the standard Feynman integrals and
dimensional regularization. Thus, the above expression reduces to
\begin{align}
\Pi _{\mu \nu }(p)=\left[-\eta _{\mu \nu }p^{2}+p_{\mu }p_{\nu }\right]\Pi
(p),
\end{align}
where the scalar polarization reads
\begin{align}
\Pi \left( p\right) =-\frac{e^{2}}{(4\pi )^{2}}\frac{1}{3}\left[\frac{2}{%
\epsilon }-\gamma \right]-\frac{e^{2}}{(4\pi )^{2}}\int_{0}^{1}dx(1-2x)^{2}%
\ln \left(\frac{4\pi \mu ^{2}}{m^{2}-x(1-x)p^{2}}\right)  \label{eqpi}
\end{align}
with $\epsilon =4-d \rightarrow 0^+$ is the ultraviolet dimensional
regularization parameter. The previous result is consistent with
relativistic covariance and the Ward-Fradkin-Takahashi identity, as in \eqref{transv}. The comparison between
\eqref{eqpi} and the known result in GSQED$_{4}$ \cite{bufalo3} leads to the conclusion that they are the same.

\subsubsection{Effective charge}

Lastly, the computation of the photon self-energy counterterm $\delta _{Z_{3}}$, vide \eqref{fotoncontrater}, which can be written directly as
\begin{equation}
\delta _{Z_{3}}=\frac{e^{2}}{3(4\pi )^{2}}\left\{ \left[\frac{2}{\epsilon }%
-\gamma \right]+\ln \left(\frac{4\pi \mu ^{2}}{m^{2}}\right)\right\}
\end{equation}%
showing explicitly that the ultraviolet divergence of the photon propagator
is absorbed by its counterterm.

It is possible now to draw some physical conclusions associated with the running of
coupling constant using as a guide the Coulomb scattering in the Born
approximation \cite{aqui,Landau}. After the renormalization procedure, the expression for the complete propagator %
\eqref{full prop
foton} can be rewritten in terms of the respective counterterm such as $(\xi =1)$
\begin{equation}
i\mathfrak{D}^{\mu \nu }\left( p\right) =\eta ^{\mu \nu }\left[ \frac{1}{%
p^{2}}-\frac{1}{p^{2}-m_{p}^{2}}\right] \left[ 1+\left[\frac{1}{p^{2}}-\frac{%
1}{p^{2} - m_{p}^{2}}\right]\left[\delta _{Z_{3}}-\Pi ^{(R)}\left( p\right) %
\right]\right].
\end{equation}
The previous relation allows a definition of the effective charge in the regime
where $k^{2}\gg m^{2}$
\begin{equation}
\alpha _{(R)}(k^{2})=\alpha (m^{2}) \left[ 1+\left[\frac{1}{p^{2}}-\frac{1}{%
p^{2} - m_{p}^{2}}\right]\left[Z_{3}-1+\frac{\alpha }{12\pi }\ln \left(\frac{%
k^{2}}{m^{2}}\right) \right]\right] ,
\end{equation}%
in which $\alpha (m^{2})=Z_3 \alpha $, and $\alpha$ is the
fine-structure constant. Besides, one can see that
\begin{equation}
\alpha _{(R)}(k^{2})=\alpha (m^{2})\left[1+\frac{\alpha (m^{2})}{12\pi }%
\frac{1}{1-\frac{k^{2}}{m_{p}^{2}}}\ln \left(\frac{k^{2}}{m^{2}}\right)%
\right]
\end{equation}
Therefore the running coupling constant expression, in the leading
logarithmic approximation, is written as follows
\begin{equation}
\frac{1}{\alpha _{(R)}(k^{2})}=\frac{1}{\alpha (m^{2})}-\frac{1}{12\pi }%
\frac{1}{1-\frac{k^{2}}{m_{p}^{2}}}\ln \left(\frac{k^{2}}{m^{2}} \right).
\label{eq 77}
\end{equation}

The expression for the running coupling constant \eqref{eq 77}
displays a pole at $k^{2}= m_{p}^{2}$; this expression then provides a
validity regime for the theory: $m^2 \ll k^{2}< m_{p}^{2}$, where the
generalized DKP theory is in fact well-defined. Moreover, this behavior is
in agreement with analysis for the fermionic and scalar theories \cite%
{bufalo2,bufalo3}.


\subsection{The DKP self-energy}

\begin{figure}
\centering
\includegraphics[width=8cm]{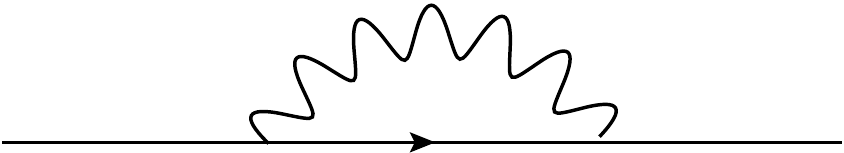}
\caption{Scalar self-energy diagram.}
\label{fig5}
\end{figure}

In the same way as in the previous case, the radiative corrections for
self-energy of the DKP particle corresponds to one diagram, as directly seen in
figure \ref{fig5}.

From the expression \eqref{sigma} rewritten in momentum representation,
\begin{equation}
\Sigma \left( p\right) =\frac{ie^{2}\beta ^{\mu }}{\left( 2\pi \right) ^{4}}%
\int d^{4}k\left[ \mathfrak{D}_{\mu \nu }\left( k\right) \mathcal{S}\left(
p-k\right) \Gamma ^{\nu }\left( p-k,p;k\right) \right] .
\end{equation}%
At the lowest order in perturbation theory with the gauge choice $\xi =1$
and the Podolsky's free parameter as $a^{2}=m_{p}^{-2}$,
\begin{equation}
\Sigma (p)=-\frac{ie^{2}m_{p}^{2}\mu ^{4-d}}{m}\int \frac{d^{d}k}{(2\pi )^{d}%
}\beta ^{\mu }[(\hat{p}-\hat{k})(\hat{p}-\hat{k}+m)-(p-k)^{2}+m^{2}]\beta
_{\mu }\frac{1}{[(p-k)^{2}-m^{2}]}\frac{1}{k^{2}(k^{2}-m_{p}^{2})}.
\end{equation}
The momentum integration is again performed using dimensional
regularization. The calculation is rather direct, and the resulting
expression is
\begin{align}
\Sigma \left( p\right) =&\frac{e^{2}m_{p}^{2}}{3m(4\pi )^{2}}
\int_{0}^{1}dx\int_{0}^{1-x}dy\frac{\beta ^{\mu }\left[ \left( 1-x\right)
^{2}\hat{p}^{2}+\left( 1-x\right) \hat{p}m-\left( 1-x\right) ^{2}p^{2}+m^{2}%
\right] \beta _{\mu } }{m^{2}x+m_{p}^{2}y -x\left(1-x\right)p^{2} }  \notag
\\
&-\frac{4e^{2}m_{p}^{2}}{3m(4\pi )^{2}}\left( 1-\beta ^{\mu }\beta _{\mu
}\right) \int_{0}^{1}dx\int_{0}^{1-x}dy\left[ \frac{2}{\epsilon }-1-\gamma
-\ln \left( \frac{m^{2}x+m_{p}^{2}y -x\left(1-x\right)p^{2}}{4\pi \mu ^{2}}%
\right) \right] .  \label{eq2a}
\end{align}

Furthermore, with help of DKP algebra (and making use of an explicit
representation, appendix \ref{sec:B}) \footnote{%
Actually, the r.h.s. of these identities is invariant under changes of
representation, showing that this result is general.} it is possible to show that $%
\beta ^{\mu }\hat{p}^{2}\beta _{\mu }=p^{2}$ and $\beta ^{\mu }\hat{p}\beta
_{\mu }=\hat{p}$. Therefore, with these identities, $\Sigma \left( p\right)$ can be conveniently separated as
\begin{equation}
\Sigma \left( p\right) = \hat{p} \Sigma _{2}\left( p^{2}\right) +\Sigma
_{1}\left( p^{2}\right) ,  \label{eq2}
\end{equation}%
in which
\begin{equation}
\Sigma _{2}\left( p^{2}\right) =\frac{e^{2}m_{p}^{2}}{3\left( 4\pi \right)
^{2}}\int_{0}^{1}dx\int_{0}^{1-x}dy\left( 1-x\right) \frac{1}{%
m^{2}x+m_{p}^{2}y-x\left( 1-x\right) p^{2}}  \label{mpdiv1}
\end{equation}%
and
\begin{align}
\Sigma _{1}\left( p^{2}\right) =& -\left( \frac{1}{\epsilon }\right) \frac{%
8e^{2}m_{p}^{2}}{3m\left( 4\pi \right) ^{2}}\left( 1-\beta ^{\mu }\beta
_{\mu }\right) +\frac{e^{2}m_{p}^{2}}{3m(4\pi )^{2}}\int_{0}^{1}dx%
\int_{0}^{1-x}dy\frac{m^{2}}{m^{2}x+m_{p}^{2}y-x\left( 1-x\right) p^{2}}%
\beta ^{\mu }\beta _{\mu }  \notag \\
& +\frac{4e^{2}m_{p}^{2}}{3m\left( 4\pi \right) ^{2}}\left( 1-\beta ^{\mu
}\beta _{\mu }\right) \int_{0}^{1}dx\int_{0}^{1-x}dy\left[ 1+\gamma +\ln
\left( \frac{m^{2}x+m_{p}^{2}y-x\left( 1-x\right) p^{2}}{4\pi \mu ^{2}}%
\right) \right] .  \label{mpdiv2}
\end{align}%
The expressions for the counter-terms of the DKP sector, $%
\delta _{Z_{0}}$ and $\delta _{Z_{1}}$, evaluated at $\alpha $-order can be presented. Then, it is time to calculate the counterterms. At first, on considering the counterterm $\delta
_{Z_{0}}$, through the relation \eqref{1cond} and equations%
\eqref{mpdiv1} and \eqref{mpdiv2}, it follows that for $\zeta =\frac{%
m_{p}^{2}}{m^{2}}>4$, it leads to
\begin{align}
-4 \delta _{Z_{0}} =&\frac{\alpha }{6\pi }\left( 4+\beta^2\right) \frac{1}{%
\epsilon _{IR}}-\frac{\alpha }{6\pi } \left( 4+ 3\beta^2 +\frac{
\Xi \beta^2}{24 }\right) -\frac{\alpha }{18\pi }\Xi^2
\left[ 36\beta^2- 180  -\zeta \left(25 \beta^2 - 109 \right)
-4\zeta ^{3}\left( 4 -\beta^2\right) \right]  \notag \\
&-\frac{\alpha }{12\pi } \left[ 4+\beta^2 +\frac{\zeta }{3}%
\left( 4(4-\beta^2 )\zeta ^{2}+3(5\beta^2-23)\zeta +48-3\beta^2\right) %
\right] \log \left[ \zeta \right]  \notag \\
&-\frac{\alpha }{12\pi } \Xi \left[ \left( \zeta \left(
2 +3\beta^2\right) -\beta^2 + 4 \right) \log \left[ \Xi -1\right] +2\left( 7
+2\beta^2 \right) \log \left[ \Xi +1\right] \right]  \notag \\
&+\frac{\alpha }{36\pi } \Xi \left[ 148  \zeta
-11\zeta ^{2}\beta^2 +4\left( 4-\beta^2 \right) \zeta ^{3}-101  \zeta
^{2}-24 -12\beta^2\right] \log \left[ \frac{\Xi - 1}{\Xi +1 }\right]  \notag \\
&-\frac{\alpha }{144\pi } \Xi\bigg[ -122\zeta \beta^2 +16\left(
4-\beta^2 \right) \zeta ^{3}- 84  -49 \beta^2  \notag \\
& +542 \zeta  -\frac{101}{3} \zeta ^{2}+\frac{23}{3}%
\zeta ^{2}\beta^2\bigg] \log \left[ \frac{\zeta -\sqrt{(\zeta -4)\zeta }-2}{\zeta +%
\sqrt{(\zeta -4)\zeta }-2}\right] ,  \label{CT1}
\end{align}
where $\beta^2 \equiv \beta _{\mu }\beta ^{\mu }$ and $\Xi
\equiv \sqrt{ \frac{\zeta }{\zeta -4}}$ have been defined and also the infrared
dimensional parameter as $\epsilon _{IR} =d-4$, $\epsilon _{IR} \rightarrow
0^-$. Similarly, the mass counterterm $\delta _{Z_{1}}$
through the relation \eqref{2cond} and equations \eqref{mpdiv1} and \eqref{mpdiv2}%
. Thus, under the condition $\zeta =\frac{m_{p}^{2}}{m^{2}}>4$ one can find that
\begin{align}
4 \delta _{Z_{1}}& =-\left( \frac{1}{\epsilon }\right) \frac{8\alpha }{3\pi }
 \left( 1-\beta^2\right) \zeta +\frac{2\alpha }{3\pi }\left( 1+\beta^2\right)
\frac{1}{\epsilon _{IR}}+\frac{ 2\alpha }{3\pi }\left( 1-\beta^2\right) \zeta %
\left[ 1+\gamma -\ln \left( \frac{4\pi \mu ^{2}}{m^{2}}\right) \right]
\notag \\
& +\frac{\alpha }{72\pi } \bigg[ 4\zeta \left( 48
-16 \zeta -9\beta^2+4\zeta \beta^2 \right) -\frac{2}{3}\zeta  \left(
4-\beta^2\right) \left( 8-18\zeta \right)  \notag \\
&-\frac{8}{3} \left( \beta^2-1\right) \left( 6\zeta ^{2}-36\zeta
-8\right) \bigg] +\frac{\alpha }{36\pi } \bigg[ %
6 \zeta \beta^2 +4 \left( 6-9\zeta +2\zeta ^{2}\right) \zeta  \notag \\
& -3\zeta \beta^2 +15\zeta ^{2}\beta^2-4\zeta ^{3}\beta^2 +12\left( \zeta +1\right) +3\beta^2
+4\left( \beta^2-1\right)\left( \zeta -6\right)\zeta ^{2} \bigg] \log %
\left[ \zeta \right]  \notag \\
& +\frac{\alpha }{36\pi }\frac{\Xi }{ \zeta } \bigg[ 3\left(
4+\beta^2 \right) \zeta \left( 5-\zeta \right) +4\left(
\beta^2-1\right) \left(4 \zeta +\zeta ^{2} -16 \right) \zeta ^{2}  \notag \\
& +4\left(4 -\beta^2\right) \left( -2\zeta ^{2}+ 13 \zeta - 20 \right)
\zeta ^{2} + 12 \left( 2- \zeta \right) \zeta \beta^2+ 3\zeta ^{3}\beta^2-12\zeta
^{2}\beta^2 +6\zeta \beta^2\bigg] \log \left[ \frac{\Xi -1 }{ \Xi + 1}\right]  \notag \\
& +\frac{\alpha }{72\pi }\frac{\Xi}{ \zeta } \bigg[ \zeta (\zeta
-4)\left( 4\left(4 -\beta^2 \right)\left(2\zeta -5 \right)\zeta +12
\right) -6\zeta \left( \zeta ^{2}-4\zeta +2\right)\beta^2  \notag \\
&+6\zeta \left( 4+\beta^2 \right) \left( 5-\zeta \right) -4 \left(\beta^2-
1\right) \left( 2 +8\zeta ^3 -32 \zeta ^{2} \right) \bigg] \log %
\left[ \frac{\zeta -\sqrt{(\zeta -4)\zeta }-2}{\zeta +\sqrt{(\zeta -4)\zeta }%
-2}\right] .  \label{CT2}
\end{align}

A pertinent comment is in place. Equation \eqref{eq2} has an UV
divergence, proportional to the $m_p ^2$-parameter. A naive thought about this divergence would present some problem with respect and spoil the
WFT identity, that yielded $Z_0=Z_2$; this is a subtle issue once the vertex part is in fact UV finite (to be treated carefully in the following). However, remarkably,
this divergence is absorbed by the mass counterterm $\delta_{Z_{1}}$, clearly at equation%
\eqref{CT2}, showing therefore that the WFT identity \eqref{wtfid} is
satisfied at this order. A similar situation was also found in the GSQED$_4$
\cite{bufalo3}.


\subsection{The vertex part}

\begin{figure}
\centering
\includegraphics[width=4.5cm]{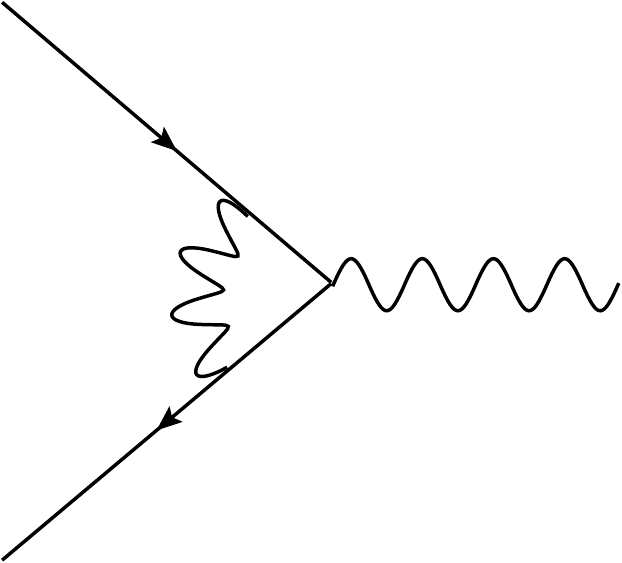}
\caption{The vertex first radiative correction.}
\label{fig6}
\end{figure}

Finally the computation of the first radiative correction associated with the
vertex function, which corresponds to the diagram depicted in figure \ref{fig6}.

The resulting outcome will be important to verify the validity of the WFT
identity as discussed in the previous subsection.

From the expression \eqref{the end}, the vertex part at the
lowest order correction is

\begin{equation}
\Lambda ^{\mu }\left( p^{\prime },p\right) =e^{2}\mu ^{4-d}\int \frac{d^{d}k
}{\left( 2\pi \right) ^{d}}\beta ^{\sigma }S\left( p^{\prime }-k\right)
\beta ^{\mu }S\left( p-k\right) \beta ^{\nu }D_{\sigma \nu }\left( k\right) ,
\end{equation}%
substituting the expressions for their respective propagators
\begin{align}
\Lambda ^{\mu }\left( p^{\prime },p\right) =& \frac{-ie^{2}\mu
^{4-d}m_{p}^{2} }{m^{2}}\int \frac{d^{d}k}{(2\pi )^{d}}\frac{ \beta ^{\sigma
}[(\hat{p}^{\prime }-\hat{k})(\hat{p}^{\prime }-\hat{k}+m)-(p^{\prime
}-k)^{2}-m^{2}]}{[(p^{\prime
}-k)^{2}-m^{2}][(p-k)^{2}-m^{2}]k^{2}\left(k^{2}-m_{p}^{2}\right)}  \notag \\
&\times \beta ^{\mu }[(\hat{p}-\hat{k})(\hat{p}-\hat{k} +m)-(p-k)^{2}-m^{2}]%
\beta _{\sigma }.
\end{align}
This expression may be simplified by making use of the Feynman
parametrization, and then be cast into a suitable form
\begin{align}
\Lambda ^{\mu }\left( p^{\prime },p\right) &=\frac{-i6e^{2}m_{p}^{2}}{m^{2}}
\mu ^{4-d}\int_{0}^{1}dx\int_{0}^{1-x}dy\int_{0}^{1-x-y}dz  \notag \\
& \times\int \frac{d^{d}k}{ \left( 2\pi \right) ^{d}} \frac{A^{\mu
}+B_{\alpha \nu }^{\mu }k^{\alpha }k^{\nu }+C_{\alpha \nu \lambda \theta
}^{\mu }k^{\alpha }k^{\nu }k^{\lambda }k^{\theta }}{\left(
k^{2}-b^{2}\right) ^{4}} ,  \label{vert}
\end{align}
in which $b^{2}=(p^{\prime }x+py)^{2}+p^{\prime
2}x+p^{2}y-m^{2}(x+y)-m_{p}^{2}z$ and the following tensor quantities
\begin{align}
A^{\mu }&= \beta ^{\sigma }\left\{ \left[ \left( 1-x\right) \hat{p} ^{\prime
}-y\hat{p}\right] \left[ \left( 1-x\right) \hat{p}^{\prime }-y\hat{p}+m%
\right] -\left[ (1-x)p^{\prime }-yp\right] ^{2}-m^{2}\right\} \beta ^{\mu }
\notag \\
& \times \left\{ \left[ \left( 1-y\right) \hat{p}-x\hat{p}^{\prime }\right] %
\left[ \left( 1-y\right) \hat{p}-x\hat{p}^{\prime }+m\right] -\left[
\left(1-y\right) p-xp^{\prime }\right] ^{2}-m^{2}\right\} \beta _{\sigma } ,
\end{align}
and
\begin{align}
B_{\alpha \nu }^{\mu }&=\beta ^{\sigma }\left[ \beta _{\alpha }\beta _{\nu
}-\eta _{\alpha \nu }\right] \beta ^{\mu }\left\{ \left[ \left( 1-y\right)
\hat{p}-x\hat{p}^{\prime }+m\right] -\left[ \left( 1-y\right) p-xp^{\prime }%
\right] ^{2}-m^{2}\right\} \beta _{\sigma }  \notag \\
&+\beta ^{\sigma }\left\{ \left[ \left( 1-x\right) \hat{p}^{\prime }-y\hat{p}%
\right] \left[ \left( 1-x\right) \hat{p}^{\prime }-y\hat{p}+m)\right] -\left[
\left( 1-x\right) p^{\prime }-yp\right] ^{2}-m^{2}\right\} \beta ^{\mu } %
\left[ \beta _{\alpha }\beta _{\nu }-\eta _{\alpha \nu }\right]  \notag \\
&-\beta ^{\sigma }\left\{ \beta _{\alpha }\left[ \left( 1-x\right) \hat{p}
^{\prime }-y\hat{p}+m)\right] +\left[ \left( 1-x\right) \hat{p}^{\prime }-y%
\hat{p}\right] \beta _{\alpha }+2\left[ \left( 1-x\right) p_{\alpha
}^{\prime }-yp_{\alpha }\right] \right\} \beta ^{\mu }  \notag \\
& \times \left\{ \beta _{\nu }[\left( 1-y\right) \hat{p}-x\hat{p}^{\prime
}+m+\left[ \left( 1-y\right) \hat{p}-x\hat{p}^{\prime }\right] \beta _{\nu
}+2\left[ \left( 1-y\right) p_{\nu }-xp_{\nu }^{\prime }\right] \right\} ,
\end{align}
and
\begin{align}
C_{\alpha \nu \lambda \theta }^{\mu }=\beta ^{\sigma }\left( \beta _{\alpha
}\beta _{\nu }-\eta _{\alpha \nu }\right) \beta ^{\mu }\left( \beta
_{\lambda }\beta _{\theta }-\eta _{\lambda \theta }\right) \beta _{\sigma }.
\end{align}
The momentum integration in expression \eqref{vert} can be evaluated and
results into
\begin{align}
\Lambda ^{\mu }(p^{\prime },p)=&\frac{e^{2}m_{p}^{2}}{(4\pi )^{2}m^{2}}
\int_{0}^{1}dx\int_{0}^{1-x}dy\int_{0}^{1-x-y}dz  \notag \\
&\times \left[\frac{A^{\mu }}{b^{4}}-\frac{g^{\alpha \nu }B_{\alpha \nu
}^{\mu }}{2b^{2}}-\frac{\Gamma (\frac{\epsilon }{2}) }{4} \left[\frac{4\pi
\mu ^{2}}{b^{2}}\right]^{\frac{\epsilon }{2}}\left(g^{\alpha \nu }g^{\lambda
\theta }+g^{\nu \theta }g^{\alpha \lambda }+g^{\theta \alpha }g^{\lambda \nu
}\right) C_{\alpha \nu \lambda \theta}^{\mu }\right] .  \label{crver}
\end{align}
The term $C_{\alpha \nu \lambda \theta }^{\mu }$ presents a
logarithmic divergence. However, by means of using the DKP algebra %
\eqref{dkpalgebra}, one can show that this term is actually vanishing its
identities%
\begin{equation}
(g^{\alpha \nu }g^{\lambda \theta }+g^{\nu \theta }g^{\alpha \lambda
}+g^{\theta \alpha }g^{\lambda \nu })C_{\alpha \nu \lambda \theta }^{\mu }=0
.  \label{div}
\end{equation}%
Showing therefore that there is no divergences on the vertex part. The
finite contribution is then given by
\begin{align}
\Lambda ^{\mu }(p^{\prime },p)=&\frac{e^{2}m_{p}^{2}}{(4\pi )^{2}m^{2}}
\int_{0}^{1}dx\int_{0}^{1-x}dy\int_{0}^{1-x-y}dz \frac{1}{\left((p^{\prime
}x+py)^{2}+p^{\prime 2}x+p^{2}y-m^{2}(x+y)-m_{p}^{2}z\right)^{2}}  \notag \\
&\times \left[\frac{A^{\mu }}{\left((p^{\prime }x+py)^{2}+p^{\prime
2}x+p^{2}y-m^{2}(x+y)-m_{p}^{2}z\right)^{2}}-\frac{g^{\alpha \nu }B_{\alpha
\nu }^{\mu }}{2}\right] .
\end{align}
This result confirm the information contained in the WTF identity \eqref{eq1}%
, $Z_0=Z_2$, assuring that the divergence of $\Sigma $ in the term
proportional to the mass $m_{p}$ \eqref{eq2} does not spoil the WTF
identity. For sake of completeness one could calculate $\delta _{Z_{0}}$ in
view of the equations \eqref{rever} and \eqref{crver}.


\section{Photon propagator and vertex at two loops}

\label{sec6}

In spite of the DKP self-energy and the photon self-energy were both
successfully renormalized, in comparison with the usual theory and also with
GQED$_4$ \cite{bufalo1,bufalo2}, the GSDKP$_4$ presents an
unexpected novel divergent structure. This novel divergence is closely
related to the one as in the self-energy in GSQED$_4$ \cite{bufalo3}. In
this way, it is important to complete the discussion of this new of
divergence ($m_P$-dependent one), present in the DKP self-energy equation %
\eqref{eq2a}, with further information and details. In particular, the diagrammatic analysis of the $\alpha ^2$-order photon
polarization tensor and the vertex function will be done in order to conclude whether
this divergence propagates and if the original counter-terms are sufficient to
control it correctly. The interest in these functions is driven mainly by
the divergence structure embedded on it from the DKP self-function.
Furthermore, discussing the photon polarization
tensor in the light of the higher-order terms, once the $\alpha $-order
calculation is not sensitive to these effects.

\begin{figure}
\includegraphics[width=16cm]{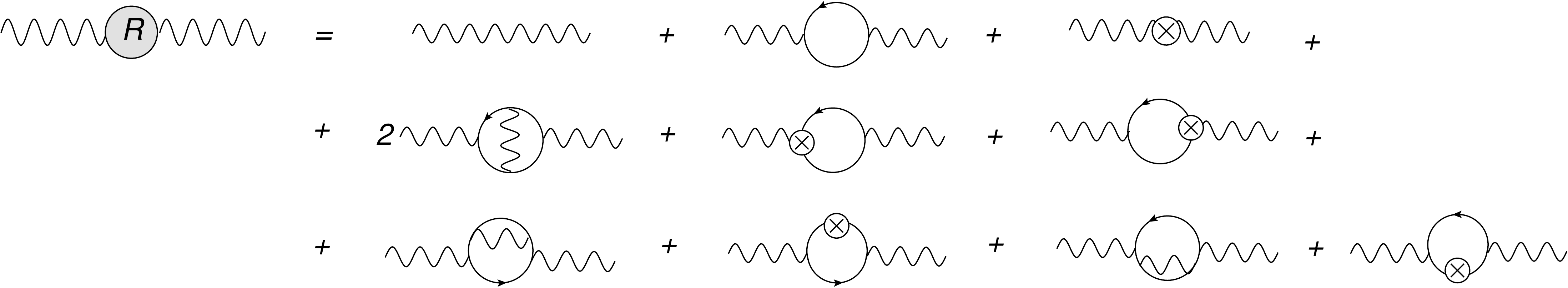}
\caption{Two-loops renormalized photon propagator.}
\label{fig7}
\end{figure}

\begin{figure}
\includegraphics[width=16cm]{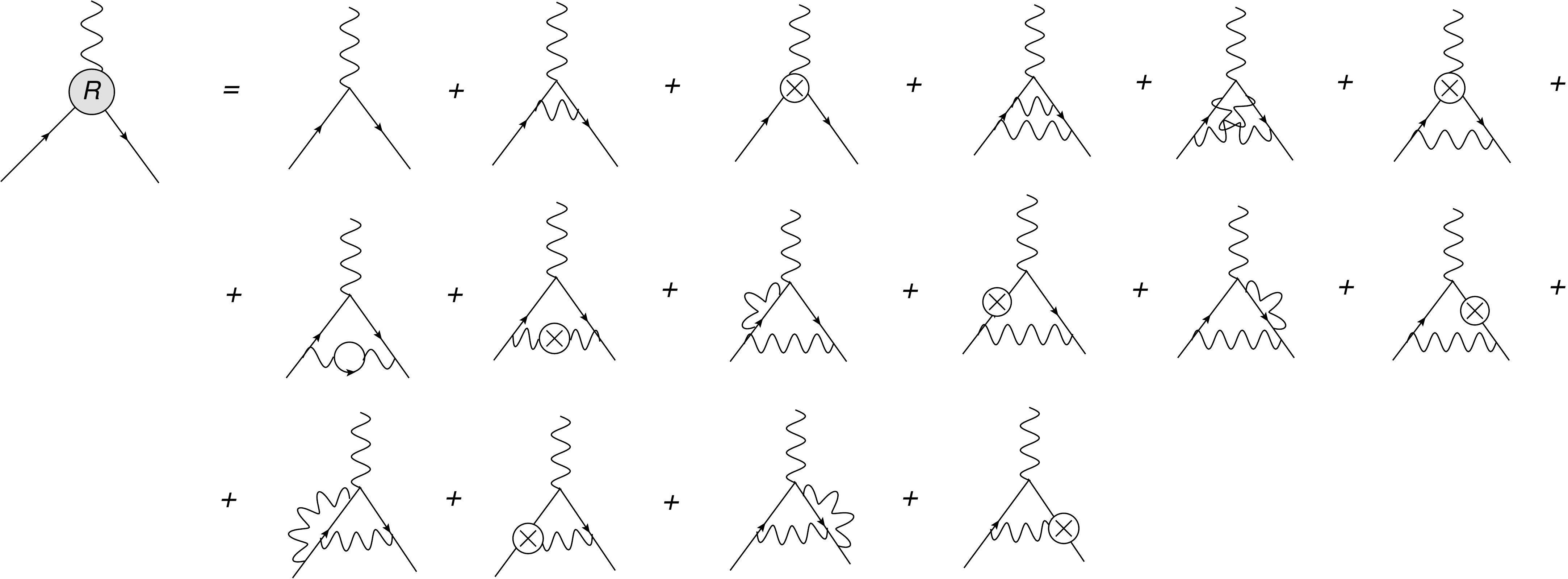}
\caption{Two-loops renormalized vertex.}
\label{fig8}
\end{figure}

The diagrams presented in this order are depicted in the figures \ref{fig7} and \ref{fig8}
for the two-loop photon polarization tensor and vertex part, respectively.

The diagrams are separated in such a way that the divergent diagram and its
respective counter-term diagram were written together. This allows to
highlight the action of the counter-terms as well as a better reading
concerning the cancellation of the $m_{p}$-dependent divergent parts.

It is worth stressing nonetheless that this work does not have the intention of
presenting here a formal proof concerning the complete renormalizability of
the theory. But the belief that a qualitative (diagrammatic) discussion does
provide all the necessary information to the
renormalizability of the theory make sense, especially regarding those $m_{p}$%
-dependent divergent diagrams. At last, this section can be concluded stating
that the original counter-terms, in particular $\delta _{Z_1}$, are
sufficient to absorb all the primitive divergences of all Green's function.


\section{Concluding remarks}

\label{sec7}

The phenomenological interaction between
scalar fields and generalized photons, from the point of view of GSDKP
electrodynamics was systematically studied. The first point to note is the implementation of the
non-mixing gauge within the Faddeev-Popov-De Witt ansatz to obtain a
covariant expression for the functional generator. Also due to the presence
of a novel divergence, the theory's renormalizability was carefully analyzed
in full detail.

The quantization of the GSDKP took place by the canonical path-integral
formalism. Based on this approach, the
Schwinger-Dyson-Fradkin equations have been derived for the basic Green's functions. In
particular, these equations provide non-perturbative information in nature of
the complete Green's functions. Besides, the SDF equations were determined
in the generalized non-mixing gauge, which, in contrast with the generalized
Lorenz condition, gave an expression for the photon's propagator in which
the transversal and the longitudinal sectors are not mixed, as one can see in
equation \eqref{final}. Along the formal development, the WFT identities have also been determined in which the gauge symmetry is proved to hold at quantum
level as well. Also, the on-shell renormalization program was applied by
including the respective counter-terms. It should be noticed that due to the
particular structure of the DKP algebra one of the renormalization
conditions in the DKP sector had to be changed in comparison with the Dirac
theory, basically because of the relation $\hat{p}\left(\hat{p}^2
-p^2\right)= 0$.

In particular, it is important to emphasize that the DKP field is often
employed in nuclear physics to describe mesons \cite{kinoshita}, when it is possible to say that it has a mesonic algebraic structure \cite{malg}. But the DKP
fields are described by the DKP algebra, while the fermionic field obeys a
Clifford algebra. Although the complete
quantum structure of the scalar field, seen by means of SDF diagrams, is exactly as those from
GQED$_{4}$ \cite{bufalo1}. In the present case, scalar DKP theory, the same diagram phenomenology for the electromagnetic interaction between
scalar or fermionic fields happens.

After concluding the formal development of the GSDKP, the evaluation of the
respective one-loop radiative corrections for the photon and DKP propagator,
and for the vertex part was done. For the DKP field self-energy and vertex part, these radiative correction expressions have a very interesting
behavior. First, it was found that the DKP field self-energy had an UV
divergence, displayed in a term proportional to $m_{p}$, as in the equation \eqref{mpdiv2}.
At first this seemed to be a problematic situation, since after evaluating
the one-loop correction to the vertex part a finite result was found, what could naively be interpreted as a violation of the the WFT identity, $Z_0 =
Z_2 $. Nonetheless, after evaluating explicitly the DKP sector counterterms,
this $m_{p}$-dependent divergence was in fact absorbed by the
mass counterterm $Z_1$. In order to verify if the renormalizability still
holds in higher-orders, i.e., whether the $m_{p}$-dependent divergence do
not propagate to higher-loops, the diagrammatic analysis of the photon self-energy and vertex part at two-loop was performed, showing that the
respective original counter-terms are sufficient to absorb all the primitive
divergences.

The information gathered from the study of the radiative
corrections of GSDKP$_{4}$ can be compared with the results for GSQED$_{4}$
\cite{bufalo3}. Then, from the self-energy of the photon one has the
same running of the effective charge (as well the validity regime for the
theory: $m^2 \ll k^{2}< m_{p}^{2}$), from the self-energy of the scalar
particle the divergence appears proportional to $m_{p}$ and the vertex has no
ultraviolet divergence. What is interesting to note is that GSQED$_{4} $ has
two vertices and GSDKP has just one, this fact makes the analysis of higher
loops apparently much easier in the framework of DKP theory rather than in
the scalar QED. Hence, based on the present outcome, it is possible to extend the present
analysis to study GSDKP at thermodynamical equilibrium within the
Matsubara-Fradkin formalism. This matter will be further elaborated and
requires deeper investigations.

\section{Acknowledgement}

R.B. thanks FAPESP for full support, T.R.C. and A.A.N. thank CAPES for full
support, and B.M.P. thanks CAPES and CNPq for partial support.

\appendix

\section{Dimensional regularization identities}

\label{sec:A}

The momentum integrals were evaluated throughout the paper by means of the
useful dimensional regularization results \cite{Framp}
\begin{align}
\int \frac{d^{d}k}{(2\pi )^{d}}\frac{1}{[k^{2}-b^{2}]^{\alpha}}=&\frac{%
i(-1)^{ \frac{d}{2}}}{(4\pi )^{\frac{d}{2}}}\frac{\Gamma (\alpha-\frac{d}{2})%
}{\Gamma (\alpha)[-b^{2}]^{\alpha-\frac{d}{2}}} ,  \label{A1} \\
\int \frac{d^{d}k}{(2\pi )^{d}}\frac{k^{\lambda }k^{\nu }}{%
[k^{2}-b^{2}]^{\alpha}}=&\frac{i(-1)^{\frac{d}{2}}}{2(4\pi )^{\frac{d}{2}}}%
\frac{\eta ^{\lambda \nu }\Gamma (\alpha-1-\frac{d}{2})}{\Gamma
(\alpha)[-b^{2}]^{\alpha-1-\frac{d}{2}}} ,  \label{A2} \\
\int \frac{d^{d}k}{(2\pi )^{d}}\frac{k^{\mu }k^{\nu }k^{\lambda
}k^{\theta }}{[k^{2}-b^{2}]^{\alpha}}=&\frac{i(-1)^{\frac{d}{2}}}{4(4\pi )^{%
\frac{d}{2}}}\frac{(\eta ^{\mu \nu }\eta ^{\lambda \theta }+\eta ^{\nu
\theta }\eta ^{\mu \lambda }+\eta ^{\theta \mu }\eta ^{\lambda \nu
})\Gamma (\alpha-2-\frac{d}{2})}{\Gamma (\alpha)[-b^{2}]^{\alpha-2-\frac{d}{2%
}}},  \label{A3}
\end{align}
in which $\eta ^{\lambda \theta }\eta _{\lambda \theta }=d$.
And the gamma's function properties
\begin{equation}
\Gamma (-n+\varepsilon )=\frac{(-1)^{n}}{n!}[\frac{1}{\varepsilon }+\Psi
_{1}(n+1)+O(\varepsilon )]  \label{A4}
\end{equation}%
in which
\begin{equation}
\Psi _{1}(n+1)=1+\frac{1}{2}+...+\frac{1}{n}-\gamma .  \label{A5}
\end{equation}%
and $\gamma $ as the Euler-Mascheroni constant. Also useful
\begin{equation}
z\Gamma (z)=\Gamma (z+1), \quad \chi ^{-\frac{\varepsilon }{2}}\simeq 1-
\frac{\varepsilon }{2}\ln \chi .  \label{A6}
\end{equation}

\section{$\protect\beta $--matrices properties}

\label{sec:B}

The $\beta $ matrices satisfy the following algebra%
\begin{equation}
\beta ^{\mu }\beta ^{\nu }\beta ^{\theta }+\beta ^{\theta }\beta ^{\nu
}\beta ^{\mu }=\beta ^{\mu }\eta ^{\nu \theta }+\beta ^{\theta }\eta ^{\nu
\mu }.  \label{B1}
\end{equation}%
A particular representation of this algebra can be given by \cite{rem}
\begin{align}
\beta ^{0} =&\left(
\begin{array}{ccccc}
0 & 0 & 0 & 0 & 1 \\
0 & 0 & 0 & 0 & 0 \\
0 & 0 & 0 & 0 & 0 \\
0 & 0 & 0 & 0 & 0 \\
1 & 0 & 0 & 0 & 0%
\end{array}%
\right) ,\quad \beta ^{1}=\left(
\begin{array}{ccccc}
0 & 0 & 0 & 0 & 0 \\
0 & 0 & 0 & 0 & 1 \\
0 & 0 & 0 & 0 & 0 \\
0 & 0 & 0 & 0 & 0 \\
0 & -1 & 0 & 0 & 0%
\end{array}%
\right) , \\
\beta ^{2} =&\left(
\begin{array}{ccccc}
0 & 0 & 0 & 0 & 0 \\
0 & 0 & 0 & 0 & 0 \\
0 & 0 & 0 & 0 & 1 \\
0 & 0 & 0 & 0 & 0 \\
0 & 0 & -1 & 0 & 0%
\end{array}%
\right) ,\quad \beta ^{3}=\left(
\begin{array}{ccccc}
0 & 0 & 0 & 0 & 0 \\
0 & 0 & 0 & 0 & 0 \\
0 & 0 & 0 & 0 & 0 \\
0 & 0 & 0 & 0 & 1 \\
0 & 0 & 0 & -1 & 0%
\end{array}%
\right) .  \label{B2}
\end{align}
The algebra \eqref{B1} can be used to show that
\begin{align}
\beta ^{\mu }\beta ^{\nu }\beta _{\mu }=\beta ^{\nu } , \quad \beta ^{\mu }%
\hat{p}^{2}\beta _{\mu }=p^{2} , \quad \beta ^{\mu }\hat{p}\beta _{\mu }=%
\hat{p}.  \label{B4}
\end{align}
In such a way that the extension of the DKP algebra over a d-dimensional spacetime leads to the following algebraic identities
\begin{equation}
\beta ^{\mu }\beta ^{\nu }\beta _{\nu }\beta _{\mu }=d , \quad \beta ^{\mu
}\beta ^{\nu }\beta _{\nu }+\beta ^{\nu }\beta _{\nu }\beta ^{\mu }=(1+d
)\beta ^{\mu }.  \label{B5}
\end{equation}

The useful property of the trace
\begin{align}
Tr\left( \beta _{\mu _{1}}\beta _{\mu _{2}}...\beta _{\mu _{2n-1}}\right)
&=0 ,  \notag \\
Tr\left( \beta _{\mu _{1}}\beta _{\mu _{2}}...\beta _{\mu _{2n}}\right)&
=\eta _{\mu _{1}\mu _{2}}\eta _{\mu _{3}\mu _{4}}...\eta _{\mu _{2n-1}\mu
_{2n}}+\eta _{\mu _{2}\mu _{3}}\eta _{\mu _{4}\mu _{5}}...\eta _{\mu
_{2n}\mu _{1}} .  \label{B3}
\end{align}


\end{document}